\documentclass[11pt]{article}
\usepackage{amsmath,amsfonts,mathtools}
\usepackage{graphicx}
\def\cN{{\cal N}}

\usepackage{ntheorem}

\makeatletter
\renewtheoremstyle{plain}
  {\item{\theorem@headerfont ##1\ ##2\theorem@separator}~}
  {\item{\theorem@headerfont ##1\ ##2\ (##3)\theorem@separator}~}
\makeatother

{\theoremheaderfont{\upshape\bfseries}
 \theorembodyfont{\normalfont\slshape}
 \newtheorem{thm}{Theorem}
 \newtheorem{prop}{Proposition}
\newtheorem{rem}{Remark}}

\usepackage[ulem=normalem,draft]{changes}
\usepackage{accents}
\usepackage{cancel}

\begin{document}
\title{On nonlinear pest/vector control via the Sterile Insect Technique: impact of residual fertility}
\author{M. Soledad Aronna$^1$\footnote{Email: soledad.aronna@fgv.br}, Yves Dumont$^{2,3,4}$ \footnote{Email: yves.dumont@cirad.fr}\\
\small{$^1$ Escola de Matem\'atica Aplicada, FGV EMAp - Rio de Janeiro, Brazil} \\
\small{$^2$CIRAD, UMR AMAP, F-97410 St Pierre, R\'eunion Island, France} \\
  \small{$^3$ AMAP, Univ Montpellier, CIRAD, CNRS, INRAE, IRD, Montpellier, France,}\\
 \small{ $^4$ University of Pretoria, Department of
   Mathematics and Applied Mathematics Pretoria, South Africa}
}

\maketitle

\begin{abstract}
We consider a minimalist model for the Sterile Insect Technique (SIT), assuming that  residual fertility can occur in the sterile male population.
Taking into account that we are able to get regular measurements from the biological system along the control duration, such as the size of the wild insect population, we study different control strategies that involve either continuous or periodic impulsive releases. We show that a combination of open-loop control with constant large releases and closed-loop nonlinear control, i.e. when releases are adjusted according to the wild population size estimates, leads to the best strategy in terms both of number of releases and  total quantity of sterile males to be released.

Last but not least, we show that SIT can be successful only if the residual fertility is less than a threshold value that depends on the wild population biological parameters. However, even for small values, the residual fertility induces the use of such large releases that SIT alone is not always reasonable from a practical point of view and thus requires to be combined with other control tools. We provide applications against a mosquito species, \textit{Aedes albopictus}, and a fruit fly, \textit{Bactrocera dorsalis}, and discuss the possibility of using SIT when residual fertility, among the sterile males, can occur.
\end{abstract}


\section{Introduction}
The Sterile Insect Technique (SIT) is a biological control technique with the advantage of targeting the pest that needs to be controlled. The concept of SIT was conceived in the 30s and 40s by three key researchers in the USSR, Tanzania and the United States (see e.g. \cite{Dyck1995} for further details about the history of SIT). The principle of SIT is very simple: it consists of releasing males that have been sterilized using ionizing radiation; these males will mate with wild females that will not produce viable offspring. However, while conceptually ``simple", SIT can be rather difficult to apply in the field as many feasibility steps need to be checked first.

Since the initial field experiments, much progress has been done under the guidance of the  IAEA (International Atomic Energy Agency), who is leading or involved in most of the SIT programs around the world. Around $30$ SIT ``feasibility" programs are currently taking place for mosquitoes. Against agricultural pests, like fruit flies, SIT programs are more advanced, such that in some places (like Spain and Mexico) effective SIT control is being practiced. Research efforts continue in order to improve SIT efficiency and also combination of SIT with other control tools (like Male Annihilation Technique). 

We believe that modeling can be an additional and efficient tool within ongoing programs in order to prevent SIT failure, improve field protocols, or test assumptions that could be difficult to verify in real conditions.

In almost all SIT models that have been studied in the last decades, the main assumption is that sterile males are $100\%$ sterile. However, in real applications this is not always true, and partial sterility has been investigated by entomologists as a possible approach to control both pests and vector populations. 
On one hand, the main drawback with full sterility is that sterile males can loose their fitness, reducing their competitiveness against wild males, such that very large, massive releases are necessary to compensate this weakness, without any warranty of success. 
On the other hand, releasing partially sterile males can be problematic as it is important to know for which level of residual fertility these releases fail to  control a wild population. This might depend on several factors, like the value of the basic offspring number, $\cN$, a threshold related to the insect population dynamics. The largest $\cN$, the more complicate it could be to efficiently control the corresponding wild population.

We build an SIT model, taking into account that the sterility induced by irradiation is not necessarily $100\%$, but can be a bit lower, such that we have a residual fertility $\epsilon$. In general, the irradiation process is made to reach $100\%$ of sterility, for a given dose of irradiation (for instance, 35-40 Gy to sterilize \emph{Aedes albopictus} pupae \cite{Oliva}). However, for some reasons (technical matters as lower dose of irradiation, environmental conditions, or others), full sterility cannot be reached. So it is important to study the impact of partial sterility on the control process.
High irradiation doses might affect the competitiveness index, $\gamma\in [0,1]$, of sterile males compared to wild males, such that we can wonder whether a lower dose, inducing residual fertility, but keeping $\gamma$ at $1$, could be an interesting strategy.

Our work stands within the framework of two SIT feasibility projects that are taking place in la R\'eunion: one against \emph{Aedes albopictus}, the TIS 2B project, funded by the French Ministry of Health and the European Regional Development Fund (ERDF); the other against a very damaging fruit fly, \textit{Bactrocera dorsalis}, that first appeared in La R\'eunion three years ago. This project, GEMDOTIS, is funded by the French government, through the EcoPhyto Call.

The SIT project against \emph{Aedes albopictus} started in 2010 after a huge epidemic of Chikungunya  impacted La R\'eunion in 2005 and in 2006. Dengue fever is also another vector-borne disease that occurs from time to time in that area with more or less virulence (the last huge dengue epidemics in La R\'eunion occurred in 1977). The dengue vector is \emph{Aedes albopictus} as well. Thus, the regional council and the French authorities decided to foster the development of biological control methods, like SIT. So far, after many years of laboratory and semi-field studies, the SIT project has started to implement local SIT releases to study the behavior and the impact of sterile males in small places. The goal for the next years is to develop releases strategies for large and focused areas, and social acceptance of the program by the local people. We believe that modeling can help to choose between different strategies and also to point out difficulties that could either drive the SIT control to failure or explain failure in field trials.

GEMDOTIS project against \textit{Bactrocera dorsalis} started in 2019. The fruit fly \textit{Bactrocera dorsalis} has been known for a long time (see \cite{Vargas2015} for an overview on {\em Bactrocera} species), but only appeared in La R\'eunion in 2017. Since then it has invaded all crops thanks to its large range of host, that is approximately $560$. However, it has some ``favorite" hosts, like guava, and mango in particular. This  is the reason why, since the arrival of this fly species, the mango production has collapsed. All biological control tools that were developed with success to lower the impact of other fruit flies, like \textit{Ceratitis rosa}, \textit{Ceratitis capitata}, and \textit{Bactrocera zonata}, are completely inefficient against \textit{Bactrocera dorsalis}. SIT is successfully used in Spain and in South Africa against \textit{Ceratitis capitata}. The objective now is to study its feasibility against \textit{Bactrocera dorsalis}, in the context of a tropical island.

In this work we consider the cases of {\em Aedes albopictus} and \textit{Bactrocera dorsalis} to illustrate our theoretical results.

The outline of the paper is as follows. In Section 2 we briefly recall the population model developed and studied in \cite{Bliman2019}, we build the partial SIT model based on continuous releases and provide conditions to control the wild population.  In Section 3  we extend the previous results to impulsive periodic releases, deriving a long term control strategy. In Section 4 we consider feedback from the models to build a closed-loop control both for continuous and periodic releases; we study different cases.  Section 5 is devoted to numerical simulations that illustrate our theoretical results.  Finally, in section 6, we summarize the main results of this paper and provide future ways to improve or extend this work.

\section{The continuous SIT model with residual fertility}

We consider the sex-structured model developed in \cite{Bliman2019}, with male $M$, and female $F$ insects, and $M_S$, the sterile males. First we assume a continuous release $\Lambda$ of sterile males. Following \cite{Barclay2001}, we assume residual fertility, i.e. there is a fraction $\epsilon$ of sterile males that remain fertile. This gives the dynamics
\begin{equation}
\quad\left\{ \begin{array}{l}
\dfrac{dM}{dt}=r\rho\dfrac{F(M+\epsilon\gamma M_{s})}{M+\gamma M_{s}}e^{-\beta(M+F)}-\mu_{M}M,\\
\dfrac{dF}{dt}=(1-r)\rho\dfrac{F(M+\epsilon\gamma M_{s})}{M+\gamma M_{s}}e^{-\beta(M+F)}-\mu_{F}F, \\
\dfrac{dM_S}{dt}=\Lambda-\mu_S M_S.
\end{array}\right.
\label{sit1}
\end{equation}
The description of the parameters is given in Table \ref{parameters} below. 
\begin{table}[h]
\centering
\begin{tabular}{ccc}
    \hline
    \textbf{Parameter} & \textbf{Description} & \textbf{Unit} \\
    \hline \hline 
    \vspace{5pt} $r$ & sex ratio & $-$ \\
    \vspace{5pt} $\rho$ & 
    \begin{tabular}{c}
        mean number of viable (that reach the adult stage)   \\
         eggs by female per day
    \end{tabular} & day$^{-1}$ \\
    \vspace{5pt} $\mu_M,\mu_F$ & male and female death rates, resp. & day$^{-1}$ \\
    \vspace{5pt} $\mu_S$ & sterile male death rate & day$^{-1}$ \\
    \vspace{5pt} $\beta$ & characteristic of the competition effect per individual & $-$ \\
    \vspace{5pt} $\gamma$ & competitiveness index of sterile male mosquitoes & $-$\\
    \vspace{5pt} $\epsilon$ & proportion of sterile males that are fertile & $-$ \\
    \hline
\end{tabular}   
\caption{Description of the parameters}
\label{parameters}
\end{table}
In general, sterilized insects have larger mortality so that 
\begin{equation}
\label{mu}
    \mu_S\geq \mu_M. 
\end{equation}
The residual fertility, $\epsilon$, of 
the sterile males is assumed to satisfy $0 \leq \epsilon<1$: when $\epsilon=0,$ this means $0$ fertility, i.e. full sterility. Similarly we have $0<\gamma \leq 1$: when $\gamma=1$, a sterile male is as competitive as wild male. In general $\gamma<1$.

\noindent Without SIT, i.e. with $M_S\equiv 0$, model \eqref{sit1} becomes
\begin{equation}
\left\{ 
\begin{split}
\dfrac{dM}{dt}&=r\rho Fe^{-\beta(M+F)}-\mu_{M}M,\\ 
\dfrac{dF}{dt}&=(1-r)\rho Fe^{-\beta(M+F)}-\mu_{F}F,
\end{split}\right.
\label{noSIT}
\end{equation}
and has been studied in \cite{Bliman2019}.

Let us consider the {\em basic offspring numbers} for the female and male populations,
\begin{equation}
    \label{basicnr}
    \mathcal{N}_{F} \coloneqq \dfrac{\left(1-r\right)\rho}{\mu_{F}},\qquad 
    \mathcal{N}_{M} \coloneqq \dfrac{r\rho}{\mu_{M}},
\end{equation} 
respectively. Then, the positive equilibrium of \eqref{noSIT} is $(M_w^*,F_w^*)$ where
\begin{equation}
\label{positive_equilibria}
M_w^* \coloneqq \dfrac{\cN_M}{\cN_F+\cN_M}\dfrac{1}{\beta} \ln \cN_F,\qquad 
F_w^* \coloneqq\dfrac{\cN_F}{\cN_F+\cN_M}\dfrac{1}{\beta} \ln \cN_F.
\end{equation}
We recall the following results (see \cite[Theorem 1]{Bliman2019}):
\begin{prop}
The following assertions hold.
\begin{itemize}
\item If $\mathcal{N}_{F}\leq1$, system (\ref{noSIT}) converges to the trivial equilibrium $\bf{0}=(0,0)$, for any non-negative initial condition.
\item If $\mathcal{N}_{F}>1$, system (\ref{noSIT}) converge to a unique positive
endemic equilibrium $(M_w^*,F_w^*)$, for any non-negative initial condition.
\end{itemize}
\end{prop}
For a matter of viability of the mosquito population in the absence of SIT, it is assume that 
$$\mathcal{N}_F>1\quad \text{and} \quad \mathcal{N}_M>1.
$$ 
Let us first assume that the release $\Lambda$ of sterile males is constant, such that, at the steady state, the number of sterile insects is 
\begin{equation}\label{Ms*}
M_{S}^* \coloneqq \dfrac{\Lambda}{\mu_{S}}.
\end{equation}
From a practical point of view, the value $M_S^*$ in \eqref{Ms*} can be reached, for instance, with massive constant releases of $2\Lambda$ during  $t=\dfrac{\ln 2}{\mu_S}$  days.
Fixing the size of the sterile population to the value $M_S^*$ given in \eqref{Ms*}, leads to the following system:
\begin{equation}
\quad\left\{ \begin{array}{l}
\dfrac{dM}{dt}=r\rho\dfrac{F(M+\epsilon\gamma M_{S}^*)}{M+\gamma M_{S}^*}e^{-\beta(M+F)}-\mu_{M}M,\\
\dfrac{dF}{dt}=(1-r)\rho\dfrac{F(M+\epsilon\gamma M_{S}^*)}{M+\gamma M_{S}^*}e^{-\beta(M+F)}-\mu_{F}F.
\end{array}\right.
\label{eq:sit_cste}
\end{equation}
Existence and uniqueness for system \eqref{eq:sit_cste} follow from standard results.

\subsection{Existence of a positive equilibrium of model (\ref{eq:sit_cste})}
Obviously ${\bf 0}=(0,0)$ is a trivial equilibrium of system (\ref{eq:sit_cste}). In order to find the positive equilibria, let us assume $M>0$ and $F>0$, and solve
\begin{equation*}
\left\{ \begin{array}{l}
r\rho\dfrac{F(M+\epsilon\gamma M_{S}^*)}{M+\gamma M_{S}^*}e^{-\beta(M+F)}=\mu_{M}M,\\
(1-r)\rho\dfrac{(M+\epsilon\gamma M_{S}^*)}{M+\gamma M_{S}^*}e^{-\beta(M+F)}=\mu_{F}.
\end{array}\right.
\end{equation*}
We get
\begin{equation}\label{eqeq}
\dfrac{(M^{*}+\epsilon\gamma M_{S}^*)}{M^{*}+\gamma M_{S}^*}e^{-\beta(M^*+F^*)}=\dfrac{1}{\mathcal{N}_{F}}\quad\text{and }\quad\dfrac{F(M^{*}+\epsilon\gamma M_{S}^*)}{M^{*}+\gamma M_{S}^*}e^{-\beta(M^*+F^*)}=\dfrac{M^*}{\mathcal{N}_{M}},
\end{equation}
where $\mathcal{N}_F$ and $\mathcal{N}_M$ were introduced in \eqref{basicnr}, so that it holds
\begin{equation}
\dfrac{F^{*}}{M^{*}}
=\dfrac{\cN_{F}}{\cN_{M}}.
\label{MF}
\end{equation}
Replacing $F^{*}$ by the latter relation in the first equation in \eqref{eqeq}
leads to
\[
\dfrac{(M^{*}+\epsilon\gamma M_{S}^*)}{M^{*}+\gamma M_{S}^*}e^{-\beta\left(1+\dfrac{\mathcal{N}_{F}}{\mathcal{N}_{M}}\right)M^{*}}=\dfrac{1}{\mathcal{N}_{F}},
\]
which is equivalent to
\[
\mathcal{N}_{F}\,e^{-\beta \left(1+\dfrac{\mathcal{N}_{F}}{\mathcal{N}_{M}}\right)M^{*}}=\dfrac{M^{*}+\gamma M_{S}^*}{M^{*}+\gamma\epsilon M_{S}^*}=1+(1-\epsilon)\dfrac{\gamma M_{S}^*}{M^{*}+\gamma\epsilon M_{S}^*}.
\]
So we aim at finding the roots of the function $f$ given by
\begin{equation}
    \label{deff}
f(x):=1+(1-\epsilon)\dfrac{a}{x+\epsilon a}-\mathcal{N}_{F}e^{-cx},
\end{equation}
where
\[
a \coloneqq \gamma M_{S}^*,\qquad 
c \coloneqq \beta  \left(1+\dfrac{\mathcal{N}_{F}}{\mathcal{N}_{M}}\right).
\]
To show the existence of a positive root of $f$ we have
to study the variation of $f$.
The case $\epsilon=0$ has been investigated in \cite{Bliman2019}. Assume
now $\epsilon>0$. In fact, we can use a similar reasoning to the one  in \cite{Bliman2019}. 

Let us first check that 
\[
f(0)=\dfrac{1}{\epsilon}-\mathcal{N}_{F}.
\]
Thus, we have three cases, $f(0)=0,$ $f(0) < 0$ and $f(0)>0$:
\begin{itemize}
\item When $\epsilon= \mathcal{N}_{F}^{-1}$, then $f(0)=0$, then there is only one non negative root, and it is $0.$
\item When $\mathcal{N}_{F}^{-1} <\epsilon <1$, then $f(0)<0$. In that case, whatever the value of $a$, $f$ admits only one positive zero. Assuming $a$, i.e. $\gamma M_S^*$, very large, we have  $f(x)\approx \dfrac{1}{\epsilon}-\mathcal{N}_{F}e^{-cx}$ which admits a positive root close to $\dfrac{\ln(\epsilon \cN_F)}{c}$. It easily follows that $\dfrac{\ln(\epsilon \cN_F)}{c}$ is a lower bound for $M^*$, this is $M^*\geq \dfrac{\ln(\epsilon \cN_F)}{c}$. It means that if partial sterility is larger than $1/\mathcal{N}_F$ then, whatever the size of the releases, the wild male population will always be greater than the positive root $M^*$ of $f$, which leads to a failure in SIT control. Of course, when $\epsilon=1$, we recover the value $M_w^*$ of the wild equilibrium, as expected.

\item When $\epsilon<\mathcal{N}_{F}^{-1}$, then $f(0)>0$. Hence, $f$
is first decreasing and then increasing such that we may have none,
one or two zeros. In fact, the number of roots might depend on $a$:
for small values of $a$, two zeros, and for large value of $a$, no zeros.
There exists $a^{\rm crit}$ such that we have only one double root
$x^{\rm crit}$ that satisfies
\[
f(x^{\rm crit})=f'(x^{\rm crit})=0.
\]
That is
\begin{equation*}
1+(1-\epsilon)\dfrac{a^{\rm crit}}{x^{\rm crit}+\epsilon a^{\rm crit}}=\mathcal{N}_{F}e^{-cx^{\rm crit}},\label{eq:1}
\end{equation*}
then
\[
(1-\epsilon)a^{\rm crit}\left(\dfrac{1}{x^{\rm crit}+\epsilon a^{\rm crit}}\right)^{2}=\mathcal{N}_{F}ce^{-cx^{\rm crit}}.
\]
Thus, putting together the two latter equalities leads to
\[
(1-\epsilon)a^{\rm crit}\left(\dfrac{1}{x^{\rm crit}+\epsilon a^{\rm crit}}\right)^{2}=c\left(1+(1-\epsilon)\dfrac{a^{\rm crit}}{x^{\rm crit}+\epsilon a^{\rm crit}}\right),
\]
which has a unique positive root $x^{\rm crit}$ given by
$$
x^{\rm crit}=\dfrac{2}{c\left(1+\sqrt{1+\dfrac{4}{a^{\rm crit}c\left(1-\epsilon\right)}}\right)}-\epsilon a^{\rm crit}.
$$
Then, replacing $x^{\rm crit}$ in (\ref{eq:1}), and setting $\Phi_{\epsilon} \coloneqq (1-\epsilon)\dfrac{a^{\rm crit}c}{2},$
\if{$$
1+(1-\epsilon)\frac{a^{\rm crit}c}{2}\left(1+\sqrt{1+2\Phi_{\epsilon}^{-1}}\right)=\mathcal{N}_{F}e^{-\epsilon ca^{\rm crit}}e^{-\dfrac{2}{1+\sqrt{1+2\Phi_{\epsilon}^{-1}}}}.$$}\fi
implies that  $\Phi_{\epsilon}$
is a positive solution of
\begin{equation}
1+\Phi_{\epsilon}\left(1+\sqrt{1+\dfrac{2}{\Phi_{\epsilon}}}\right)
=\mathcal{N}_{F}e^{-2\dfrac{\epsilon}{1-\epsilon}\Phi_{\epsilon}}
e^{-\dfrac{2}{1+\sqrt{1+\frac{2}{\Phi_{\epsilon}}}}}.\label{eq:TransEq}
\end{equation}
\end{itemize}
Summarizing, we get the result below.
\begin{prop}
\label{prop1}
The following assertions hold.
\begin{itemize}
    \item[(i)] If $\epsilon<\mathcal{N}_{F}^{-1}$, then
there exists $\Lambda_{\rm crit}^{\epsilon}>0$ such that system \eqref{eq:sit_cste}
admits
\begin{itemize}
\item two positive distinct equilibria if $0<\Lambda<\Lambda_{\rm crit}^{\epsilon}$,
\item one positive equilibrium if $\Lambda=\Lambda_{\rm crit}^{\epsilon}$,
\item no positive equilibria if $\Lambda>\Lambda_{\rm crit}^{\epsilon}$.
\end{itemize}
The value of $\Lambda_{\rm crit}^{\epsilon}$ is defined by 
\[
\Lambda_{\rm crit}^{\epsilon}\coloneqq \dfrac{2}{1-\epsilon}\dfrac{\mu_{S}}{\beta\gamma\left(1+\frac{\mathcal{N}_{F}}{\mathcal{N}_{M}}\right)}\Phi_{\epsilon},
\]
where $\Phi_{\epsilon}$ is the unique positive solution to the transcendental
equation (\ref{eq:TransEq}).
\item[(ii)] Assume that $\mathcal{N}_{F}^{-1}<\epsilon <1$, then, for any $\Lambda>0$, i.e. for any $M_S^*>0$, the system \eqref{eq:sit_cste} admits one positive equilibrium, bounded from below (component-wise) by the point
\begin{equation}
(M^*_\ell,F^*_\ell) \coloneqq \frac{\ln (\epsilon\cN_F)}{\beta(\cN_F+\cN_M)}\big(\cN_M,\cN_F\big).
\label{lower_bound}
\end{equation}
\end{itemize}
\end{prop}


Clearly, if partial sterility is too large, SIT will fail: even very large releases will only have a small effect on the wild population.

\subsection{Asymptotic analysis of the equilibria}

We assume $\Lambda > \Lambda_{\rm crit}^{\epsilon}$ such that system \eqref{eq:sit_cste} possesses only the trivial equilibrium (as established in Proposition \ref{prop1}). We compute the Jacobian related to system  \eqref{eq:sit_cste}, that gives $J(M,F)$ equals to
{\small\[
\left(\begin{array}{cc}
r\rho F \Delta \left(\dfrac{(1-\epsilon)\gamma M_{S}^*}{M+\gamma M_{S}^*}-\beta(M+\epsilon\gamma M_{S}^*)\right) -\mu_M
& r\rho(M+\epsilon\gamma M_{S}^*)(1-\beta F)\Delta \\
(1-r)\rho F \Delta \left(\dfrac{(1-\epsilon)\gamma M_{S}^*}{M+\gamma M_{S}^*}-\beta(M+\epsilon\gamma M_{S}^*)\right) & (1-r)\rho(M+\epsilon\gamma M_{S}^*)(1-\beta F)\Delta-\mu_{F}
\end{array}\right),
\]}
where $\Delta \coloneqq 
\dfrac{e^{-\beta(M+F)}}{M+\gamma M_{S}^*}.$
Computing $J$ at ${\bf 0}$ gives
\[
J(0,0)=\left(\begin{array}{cc}
-\mu_{M} & r\rho\epsilon\\
0 & (1-r)\rho\epsilon-\mu_{F}
\end{array}\right).
\]
Thus, if $\epsilon<{\mathcal{N}_{F}^{-1}}$, then $\bf{0}$ is
Locally Asymptotically Stable (LAS). Otherwise it is unstable. The condition $\epsilon<\mathcal{N}_{F}^{-1}$ is also necessary in order to guarantee that the population can be controlled and become as small as necessary in a finite ``short'' time.



To show that $\bf{0}$ is Globally Asymptotically Stable (GAS), like in \cite{Bliman2019}, we use the Dulac criterion, thanks to the following Dulac function:
\[
\psi(M,F) \coloneqq \dfrac{M+\gamma M_{S}^*}{F(M+\epsilon\gamma M_{S}^*)}.
\]
We compute
\begin{multline*}
    \dfrac{\partial}{\partial M}\left(r\rho e^{-\beta(M+F)}-\mu_{M}\dfrac{M+\gamma M_{S}^*}{F(M+\epsilon\gamma M_{S}^*)}M\right)=\\-r\rho\beta e^{-\beta(M+F)}-\dfrac{\mu_{M}}{F}\left(\dfrac{\epsilon\gamma M_{S}^*(2M+\gamma M_{S}^*)+M^{2}}{(M+\epsilon\gamma M_{S}^*)}\right)<0,
\end{multline*}
and
\[
\dfrac{\partial}{\partial F}\left((1-r)\rho e^{-\beta(M+F)}-\mu_{F}\dfrac{M+\gamma M_{S}^*}{(M+\epsilon\gamma M_{S}^*)}\right)=-(1-r)\beta\rho e^{-\beta(M+F)}<0.
\]
Thus, according to Poincar\'e-Bendixson Theorem, since $\bf{0}$ is the only LAS equilibrium when $\epsilon<\mathcal{N}_{F}^{-1}$ and the system has no closed orbits, we deduce that $\bf{0}$ is also GAS.

Assume now that $0<\Lambda <\Lambda_{\rm crit}^{\epsilon}$. Then, according to Proposition \ref{prop1}, there exists two positive equilibria, that we call $\bf{E}_1^*$ and $\bf{E}_2^*$, with $\bf{E}_1^*$ is unstable and $\bf{E}_2^*$  LAS. Like for the case $\epsilon=0$ (see \cite{Bliman2019}), we have bistability, and the basin of attraction of $\bf{0}$ contains the interval 
\[
[{\bf 0},{\bf E_1}^*[\coloneqq\{(M,F)\in \mathbb{R}^2_+: 0\leq M<M_1^*, 0\leq F < F_1^*\},
\] 
and the basin of attraction of ${\bf E_2}^*$ contains the interval 
\[
]{\bf{E^*_1}},+\infty [ \coloneqq\{(M,F)\in \mathbb{R}^2_+: M_1^*<M, F_1^*<F \}.
\]

\section{Impulsive periodic releases}
To achieve practicable strategies, we consider impulsive periodic releases, since the releases in the field are done instantaneously and periodically, i.e $M_S$ follows a dynamics of the form
\begin{equation}
\left\{
\begin{array}{l}
\dfrac{dM_S}{dt}=-\mu_S M_S, \quad \mbox{for  $t\neq n\tau$,}\\
M_S(t_c+n\tau^+)=M_S(t_c+n\tau)+\tau \Lambda, \qquad n\in \mathbb{N},
\end{array}
\right.
\end{equation}
where $t_c$ is the time at which the control starts.
We assume that releases are done every $\tau>0$ days, such that the number of sterile males asymptotically approaches the function $M_{s,{\rm per}}$ given by (see \cite{Bliman2019}):
\[
M_{s,{\rm per}}(t)\coloneqq\dfrac{\tau\Lambda}{1-e^{-\mu_{S}\tau}}e^{-\mu_{S}\left(t-\left\lfloor\frac{t}{\tau}\right\rfloor \tau\right)}.
\]
Thus we derive the following system with periodic coefficients
\begin{equation}
\left\{ \begin{array}{l}
\dfrac{dM}{dt}=r\rho\dfrac{F(M+\epsilon\gamma M_{s,{\rm per}})}{M+\gamma M_{s,{\rm per}}}e^{-\beta(M+F)}-\mu_{M}M,\\
\dfrac{dF}{dt}=(1-r)\rho\dfrac{F(M+\epsilon\gamma M_{s,{\rm per}})}{M+\gamma M_{s,{\rm per}}}e^{-\beta(M+F)}-\mu_{F}F.
\end{array}\right.\label{eq:periodicTIS}
\end{equation}
The pest/vector free equilibrium is still an equilibrium of system (\ref{eq:periodicTIS}). Like in the previous part, the objective is to find conditions under which the equilibrium ${\bf 0}$ is GAS for system (\ref{eq:periodicTIS}).
From (\ref{eq:periodicTIS}), we have
\begin{equation}
\dfrac{dF}{dt}=\left((1-r)\rho\dfrac{M+\epsilon\gamma M_{s,{\rm per}}}{M+\gamma M_{s,{\rm per}}}e^{-\beta(M+F)}-\mu_{F}\right)F.\label{eq:F_TIS}
\end{equation}
Thus,
\[
\dfrac{M+\epsilon\gamma M_{s,{\rm per}}}{M+\gamma M_{s,{\rm per}}}e^{-\beta(M+F)}=\left(\dfrac{(1-\epsilon)M}{M+\gamma M_{s,{\rm per}}}+\epsilon\right)e^{-\beta(M+F)}\leq\dfrac{(1-\epsilon)\alpha}{\gamma M_{s,{\rm per}}}+\epsilon,
\]
where $\alpha \coloneqq \max\{xe^{-\beta x}: x\geq 0\}=\dfrac{1}{e\beta}$ (see \cite{Bliman2019}). Then, integrating (\ref{eq:F_TIS})
between $n\tau$ and $t \geq n\tau$, we derive
\[
F(t)\leq F(n\tau) \exp\int_{n\tau}^{t}\left((1-r)\rho\left(\dfrac{(1-\epsilon)\alpha}{\gamma M_{s,{\rm per}}}+\epsilon\right)-\mu_{F}\right)ds.
\]
Taking $t=(n+1)\tau$, we deduce
\[
F((n+1)\tau)\leq F(n\tau) \exp\left((1-r)\rho \tau\left[\dfrac{(1-\epsilon)\alpha}{\gamma}\left\langle \dfrac{1}{M_{s,{\rm per}}}\right\rangle -\left(\mathcal{N}_F^{-1}-\epsilon \right)\right]\right),
\]
where $\displaystyle\left\langle \dfrac{1}{M_{s,{\rm per}}}\right\rangle\coloneqq \frac{1}{\tau}\int_0^\tau \dfrac{1}{M_{s,{\rm per}}(t)}dt$.
Therefore, since $\epsilon<\mathcal{N}_F^{-1}$, the sequence $\left\{ F(n\tau)\right\} _{n\in\mathbb{N}}$ decreases towards $0$, if 
\[
\dfrac{(1-\epsilon)\alpha}{\gamma}\left\langle \dfrac{1}{M_{s,{\rm per}}}\right\rangle -\left(\mathcal{N}_F^{-1}-\epsilon\right)<0,
\]
that is 
\begin{equation}\label{Lambdacond}
\dfrac{2\left(\cosh\left(\mu_{s}\tau\right)-1\right)}{\mu_{S}\tau^{2}\Lambda}<\dfrac{\gamma}{(1-\epsilon)\alpha}\left(\mathcal{N}_F^{-1}-\epsilon\right),
\end{equation}
since $\displaystyle\left\langle \dfrac{1}{M_{s,{\rm per}}}\right\rangle = \dfrac{2\left(\cosh\left(\mu_{s}\tau\right)-1\right)}{\mu_{S}\tau^{2}\Lambda}$ (see \cite{Bliman2019}). Inequality \eqref{Lambdacond} holds if
\begin{equation}
\Lambda>\Lambda_{\rm crit}^{per} \coloneqq \dfrac{2\left(\cosh\left(\mu_{s}\tau\right)-1\right)}{\mu_{S}\tau^{2}}\dfrac{(1-\epsilon)\mathcal{N}_F}{\gamma\left(1-\epsilon\mathcal{N}_F\right)e\beta}
\label{eq:cond}
\end{equation}
This is sufficient to ensure that $F$ converges towards $0$,
which induces the same behavior for $M$. Thus,  condition \eqref{eq:cond}
implies that ${\bf 0}$ is also GAS. We derive the following result.

\begin{thm}
For any given $\tau>0$, assuming that $\Lambda$ is chosen such that (\ref{eq:cond}) is verified, every solution of system (\ref{eq:periodicTIS}) converges to ${\bf 0}$.
\end{thm}

Thus, using Theorem 1, massive releases, i.e. $\tau\Lambda = k \times \tau \left(\lfloor \Lambda_{crit}^{per}\rfloor+1\right)$ with $k\geq 1$, guarantee that the system will be driven close to zero in finite time. However, once the control stops, the system will recover and the population will reach their initial (positive) equilibrium. Also, for real applications, massive releases are not sustainable and can only be conducted for a limited time. Once the system is closed to zero, small releases would be preferable in order to maintain the wild population at a low level (which can be determined evaluating the epidemiological risk and/or an economical threshold value). We follow the same strategy developed in \cite{Yatat2020,Tapi2020}.

\subsection{Long term control strategy for periodic releases}\label{SSbox}
System \eqref{eq:periodicTIS} can be bounded from above by the following system
\begin{equation}
\left\{ \begin{array}{l}
\dfrac{dM}{dt}=r\rho\dfrac{F(M+\epsilon\gamma M_{s,l})}{M+\gamma M_{s,l}}e^{-\beta F}-\mu_{M}M,\\
\dfrac{dF}{dt}=(1-r)\rho\dfrac{F(M+\epsilon\gamma M_{s,l})}{M+\gamma M_{s,l}}e^{-\beta F}-\mu_{F}F,
\end{array}\right.
\label{monotone}    
\end{equation}
where $M_{s,l}>0$ is a lower bound of $M_{s,{\rm per}}(t)$ given by:
is
\[
M_{s,l}\coloneqq \dfrac{\tau\Lambda}{1-e^{-\mu_{S}\tau}}e^{-\mu_{S}\tau}.
\]
In fact it is easy to check that system (\ref{monotone}) is a monotone cooperative system within the subset $\mathcal{S} \coloneqq\left\{ (F,M)\in\mathbb{R}_{+}^{2}: F<\dfrac{1}{\beta}\right\} $.
Hence, once the solution of the periodic system (\ref{eq:periodicTIS}), after several ``massive"
releases, enters $\mathcal{S}$, we can use the fact that, for a given (small)
release $M_{s,{\rm obj}}$, the equilibria $\bf{0}$, $\bf{E}_{1}$, and $\bf{E}_{2}$, of system (\ref{monotone}), are ordered, i.e. $\bf{0}<E_1<E_2$. In particular, the box $[{\bf 0},\bf{E}_{1}[$ is included in the basin of attraction of ${\bf 0}$. 

This last result allows us to deduce a long term control strategy that can be split in two phases: a first initial finite phase with massive releases (where ${\bf 0}$ is GAS) to enter $[{\bf 0},\bf{E}_{1}[$; followed by a second infinite phase, where control is insured by small releases.

The first phase is finite in time, meaning that there exists a time $t^*>0$, such that for all $t>t^*$, $(M(t),F(t)) \in [{\bf 0},\bf{E}_{1}[$. The existence and an upper bound of $t^*$ can be estimated using the same approach in \cite{Strugarek2019}.

Practically, for a given small release $M_{s,{\rm obj}}$, we have to estimate $\bf{E}_{1}$. This is done by finding the zeros of the function $f$ in \eqref{deff} with
\[
x=F,\qquad a=\gamma M_{s,l},\qquad c=\beta\left(1+\dfrac{\mathcal{N}_{M}}{\mathcal{N}_{F}}\right).
\]
It suffices to estimate the
smallest positive root of $f$ to derive $F_{1}$, then (analogously to system \eqref{eq:sit_cste}), we have $M_{1}=\dfrac{\mathcal{N}_{M}}{\mathcal{N}_{F}}F_{1}$.

\section{Closed-loop control approach}
In the previous control approach, for the continuous and periodic cases, we did not consider information on the system along the control duration: the size of the releases was only related to the initial value of the population, at the wild equilibrium. In general, several tools exist that may provide information on the wild population size along the year and during the control, such that it is of interest to take into account this information in order to adapt the size of the releases. This is what is done when using a closed-loop control approach.

Here we let $\kappa \colon [0,+\infty) \to \mathbb{R}_0^+$ be a function such that
 $M_{s}(t)={\boldsymbol\kappa}(t)M(t).$ Then (\ref{eq:sit_cste}) becomes
\begin{equation}
\label{model7bis}
\left\{
\begin{split}
\dfrac{dM}{dt} &=r\rho F\dfrac{(1+\epsilon\gamma {\boldsymbol\kappa})}{1+\gamma {\boldsymbol\kappa}}e^{-\beta(M+F)}-\mu_{M}M,\\
\dfrac{dF}{dt}&=(1-r)\rho\left(\dfrac{1+\epsilon\gamma {\boldsymbol\kappa}}{1+\gamma {\boldsymbol\kappa}}e^{-\beta(M+F)}-\cN_F^{-1}\right)F.
\end{split}
\right.
\end{equation}
Let us impose that there exists $\theta>0$ such that
\begin{equation}
\dfrac{1+\epsilon\gamma {\boldsymbol\kappa}}{1+\gamma {\boldsymbol\kappa}}e^{-\beta(M+F)}-
\cN_F^{-1}\leq -\theta,
\label{12}
\end{equation}
which is equivalent to choosing $\boldsymbol\kappa$ such that
\begin{equation}
{\boldsymbol\kappa}(t) \geq \frac1\gamma\frac{e^{-\beta(M(t)+F(t))}-(\cN_F^{-1}-\theta)}{(\cN_F^{-1}-\theta)-\epsilon e^{-\beta(M(t)+F(t))}}.
\label{13}
\end{equation}
Note also that (\ref{12}) only makes sense if $\theta\leq  \cN_F^{-1}$. In order to always have positive and finite values in the r.h.s. term of (\ref{13}), the following condition is needed
\[
\epsilon+\theta <\cN_F^{-1}.
\]
Then, from \eqref{model7bis} and \eqref{12}, we deduce that $ \dfrac{dF}{dt} \leq -(1-r)\rho\theta F(t)$ which implies 
\begin{equation}
\label{estimateF}
F(t) \leq F(0)e^{-(1-r)\rho\theta t}.
\end{equation}
This yields that $F$ converges exponentially to $0$ when $t$ goes to $+\infty$. Then we deduce that
\begin{equation*}
\begin{split}
\dfrac{dM}{dt}(t) &\leq r\rho F(0)e^{-(1-r)\rho\theta t}\left(\cN_F^{-1} -\theta\right) -\mu_M M.
\end{split}
\end{equation*}
Applying Gronwall's Lemma to latter inequality leads to
\begin{equation}
\label{estimateM}
M(t) \leq M(0)e^{-\mu_M t}+F(0) \frac{r\rho(\mathcal{N}_F^{-1}-\theta) }{\mu_M-(1-r)\rho\theta} \left(e^{-(1-r)\rho\theta t}-e^{-\mu_M t}\right),
\end{equation}
so that $M$ also converges exponentially to $0$ when $t$ goes to $+\infty$.

From the previous computations, we deduce the following result.
\begin{prop}[Continuous nonlinear feedback release] 
\label{prop2} 
For a given nonnegative $\epsilon< \cN_F^{-1}$, let $\theta$ be a positive real number such that
\begin{equation}
0<\theta+\epsilon<\cN_F^{-1}
\label{eq:1-1}.
\end{equation}
If $M_S$ is chosen such that
\[
M_S(t) \geq \kappa\big(M(t)+F(t)\big) M(t),
\]
where 
\begin{equation}
\displaystyle \kappa(x) :=  \frac1\gamma\frac{e^{-\beta x}-(\mathcal{N}_F^{-1}-\theta)}{(\mathcal{N}_F^{-1}-\theta)-\epsilon e^{-\beta x}},
\label{kappa}
\end{equation}
then every solution of \eqref{eq:sit_cste} converges exponentially to
${\bf 0}$.
\end{prop}

\begin{rem}
The function $\kappa $ in \eqref{kappa} gives a {\em nonlinear feedback law} for $M_S.$
The continuous linear feedback control result obtained in \cite{Bliman2019} can be recovered by considering an upper bound for $\boldsymbol\kappa(t)$, obtained when $M+F=0$, and setting $k \coloneqq \cN_F^{-1}-\theta$, that gives
\begin{equation}
    \label{linearfeedback}
{\boldsymbol\kappa}(t) =\kappa(0)= \frac1\gamma\frac{1-k}{k-\epsilon}
\end{equation}
\end{rem}


\begin{rem}[On the choice of $\theta$]
\label{rem2}
Note that, in view of \eqref{eq:1-1}, one has that $\theta=r_1 \cN_F^{-1}$ and $\epsilon=r_2 \cN_F^{-1},$ with $r_1\in (0,1), r_2\in [0,1)$ and $r_1+r_2<1.$ Simple calculations lead to the following alternative expression for the feedback law $\kappa:$
\[
\kappa(x) = \frac{1}{\gamma} \frac{e^{-\beta x}(\cN_F^{-1}-r_2)}{1-r_1-r_2 e^{-\beta x}}-1
\]
So that, for a fixed value of $\epsilon$ (i.e. of $r_2$),  the gain $\kappa$ increase w.r.t. to $\theta$ (i.e. w.r.t. $r_1$). The same happens to  the speed of convergence of $F$ to 0, that is proportional to $\theta.$
\\
In particular, when $\epsilon$ is close to $\cN_F^{-1}$, i.e. $r_2$ close to 1, then $\theta$ is close to 0 so that the convergence of $F$ to 0 is slow but, at the same time, the size of the gain $\kappa$ is small.
\end{rem}

\subsection{Impulsive releases - synchronized measurements and releases}
Let us now consider that we release sterile insects with a period of $\tau.$ 
From \eqref{estimateF} and \eqref{estimateM} we get, for $t\in [n\tau,(n+1)\tau),$
\begin{equation*}
\begin{split}
F(t) &\leq F(n\tau) e^{-(1-r)\rho\theta (t-n\tau)},\\
M(t) &\leq M(n\tau) e^{-\mu_M (t-n\tau)}   + F(n\tau) \frac{r\rho(\mathcal{N}_F^{-1}-\theta) }{\mu_M-(1-r)\rho\theta} \left(e^{-(1-r)\rho\theta (t-n\tau)}-e^{-\mu_M (t-n\tau)}\right).
\end{split}
\end{equation*}
We impose the condition
\begin{equation}
M_S(t) \geq {\boldsymbol\kappa}(t) M(t),\quad t\in [n\tau,(n+1)\tau).
\end{equation}
This is verified if, for $t\in [n\tau,(n+1)\tau),$
\begin{multline*}
    M_S(t) \geq  {\boldsymbol\kappa}(t) \left(M(n\tau) e^{-\mu_M (t-n\tau)}  + F(n\tau)\frac{r\rho(\mathcal{N}_F^{-1}-\theta) }{\mu_M-(1-r)\rho\theta} \left(e^{-(1-r)\rho\theta (t-n\tau)}-e^{-\mu_M (t-n\tau)}\right)\right).
\end{multline*}
Since $\kappa$, introduced in \eqref{kappa}, decreases as a function of $M+F,$ and $M$ and $F$ remain larger than $M(n\tau)e^{\mu_M (t-n\tau)}$ and $F(n\tau)e^{\mu_F (t-n\tau)},$ respectively, we get that
\begin{equation*}
{\boldsymbol\kappa}(t) =\kappa(M(t)+F(t)) \leq \kappa\left(M(n\tau)e^{\mu_M (t-n\tau)}+F(n\tau)e^{\mu_F (t-n\tau)} \right) =: \kappa_{\max}^n.
\end{equation*}
Thus, if for $s\in [0,\tau),$ it holds
\begin{multline}
  \label{MSineq}
M_S(n\tau+s) = \big(M_S(n\tau)+\tau\Lambda_n\big) e^{-\mu_Ss} \geq \\  \kappa_{\max}^n \Big( M(n\tau) e^{-\mu_M s}   + F(n\tau)\frac{r\rho(\mathcal{N}_F^{-1}-\theta) }{\mu_M-(1-r)\rho\theta}\left(e^{-(1-r)\rho\theta s}-e^{-\mu_M s}\right)\Big).
\end{multline}
then $(M,F)$ converges asymptotically to 0. This last equation is equivalent to
\begin{multline*}
  \tau\Lambda_n \geq  -M_S(n\tau) 
+ \kappa_{\max}^n \Big(M(n\tau) e^{(\mu_S-\mu_M)s}   
+ F(n\tau) \frac{r\rho(\mathcal{N}_F^{-1}-\theta) }{\mu_M-(1-r)\rho\theta} \left(e^{(\mu_S-(1-r)\rho\theta) s}-e^{(\mu_S-\mu_M) s}\right)\Big),
\end{multline*}
Since $\mu_S \geq \mu_M$ and $\theta \leq \cN_F^{-1}$, assuming the additional condition $\theta \leq \dfrac{\mu_S}{\mu_F} \cN_F^{-1}$, one has that all the coefficients and exponents in the r.h.s. of latter expression are positive, so a stronger inequality is obtained if we take $s=\tau$ for the exponential expressions with positive coefficient and $s=0$ for the one with negative coefficient. This is, we impose 
\begin{multline*}
\tau\Lambda_n \geq  -M_S(n\tau) + \kappa_{\max}^n e^{(\mu_S-\mu_M)\tau} \Big(M(n\tau)  + F(n\tau) \frac{r\rho(\mathcal{N}_F^{-1}-\theta) }{\mu_M-(1-r)\rho\theta} \left(e^{({\mu_M}-(1-r)\rho\theta) \tau}-1\right)\Big).
\end{multline*}
We summarize the result as follows.

\begin{thm}[Sufficient condition for stabilization by  impulsive feedback control] \label{th1} 
For a given non negative $\epsilon$ and a positive $\theta$ such that $\theta+\epsilon < \min\left(1,\dfrac{\mu_S}{\mu_F}\right)\mathcal{N}_{F}^{-1}$,
assume that for any $n\in\mathbb{N}$
\[
\tau\Lambda_{n}\ \geq\left|K_{\epsilon,n}\begin{pmatrix}
M(n\tau)\\
F(n\tau)
\end{pmatrix}-M_{S}(n\tau)\right|_{+},
\]
with
{\small
\begin{equation*}
    \begin{split}
        K_{\epsilon,n}:= \frac{1}{\gamma} \dfrac{e^{-\beta(M_{n+1}+F_{n+1})}-\left(\mathcal{N}_{F}^{-1}-\theta\right)}{\left(\mathcal{N}_{F}^{-1}-\theta\right)-\epsilon e^{-\beta(M_{n+1}+F_{n+1})}}e^{(\mu_{S}-\mu_{M})\tau}\left(1 , \frac{r\rho(\mathcal{N}_{F}^{-1}-\theta)}{\mu_{M}+(1-r)\rho\theta}\left(e^{\left(\mu_{M}-(1-r)\rho\theta\right)\tau}-1\right)\right), 
    \end{split}
\end{equation*}
}
where 
\[
M_{n+1}\coloneqq M(n\tau)e^{-\mu_{M}\tau}\qquad\mbox{and}\qquad F_{n+1}\coloneqq F(n\tau)e^{-\mu_{F}\tau}.
\]
Then, every solution of system \eqref{eq:sit_cste} converges exponentially
towards $\mathbf{0}$, with a convergence rate bounded from below
by a value independent of the initial condition.

If, moreover 

\[
\tau\Lambda_{n}\leq K_{\epsilon,n}\begin{pmatrix}
M(n\tau)\\
F(n\tau)
\end{pmatrix},
\]
then the series of impulses $\sum\limits _{n=0}^{+\infty}\Lambda_{n}$
converges. 
\end{thm}

\begin{rem}
We recover the impulsive linear feedback control of \cite{Bliman2019} when we replace $K_{\epsilon,n}$ in previous theorem with
{\small\[
\dfrac{1}{\gamma} \dfrac{1-\left(\mathcal{N}_{F}^{-1}-\theta\right)}{\left(\mathcal{N}_{F}^{-1}-\theta\right)-\epsilon}e^{(\mu_{S}-\mu_{M})\tau}\begin{pmatrix}1, \dfrac{r\rho}{\mu_{M}+(1-r)\rho\theta}\left(\dfrac{1}{\mathcal{N}_{F}}-\theta\right)\left(e^{\left(\mu_{M}-(1-r)\rho\theta\right)\tau}-1\right)\end{pmatrix}.
\]}
\end{rem}

\subsection{Sparse measurements}
It is reasonable to expect that measurements of the size of the wild female and male populations are not done very frequently. Having this in mind, we assume in this part that measurements are done every $p\tau$ days, with $p\in \mathbb{N}^*$.
Like in \cite{Bliman2019}, we need to adapt the proof of previous Theorem \ref{th1}.

We have, for $m=0,\dots,p-1,$ and $s\in [0,\tau),$
\begin{equation*}
\begin{split}
&M_S(s+(np+m)\tau) = \big(\Lambda_{np+m}\tau + M_S((np+m)\tau) \big) e^{-\mu_S s} \\
&= \Big(\Lambda_{np+m} \tau + \Lambda_{np+m-1} \tau e^{-\mu_S \tau} + \dots + \Lambda_{np} \tau e^{-m\mu_S\tau} + M_S(np\tau)e^{-m\mu_S\tau} \Big)e^{-\mu_S s}.
\end{split}
\end{equation*} 
We have,
\begin{equation*}
\kappa\big( (np+m)\tau + s \big) \leq \kappa\big( M_{\min}^{np+m}+F_{\min}^{np+m}\big)=: \kappa_{\max}^{np+m},
\end{equation*}
where 
\begin{equation}
\label{MFmin}
M_{\min}^{np+m} := M(np\tau) e^{-m\mu_M \tau},\qquad 
F_{\min}^{np+m} := F(np\tau) e^{-m\mu_F \tau}.
\end{equation}
As done above in \eqref{MSineq}, we impose
\begin{equation*}\begin{split}
  \Big(\Lambda_{np+m} \tau &+ \Lambda_{np+m-1} \tau e^{-\mu_S \tau} + \dots + \Lambda_{np} \tau e^{-m\mu_S\tau} + M_S(np\tau)e^{-m\mu_S\tau} \Big)e^{-\mu_S s}  \\
&\geq \kappa_{\max}^{np+m}\Big( M(np\tau) e^{-\mu_M(m\tau+s)} \\ &+ F(np\tau) \frac{r\rho(\mathcal{N}_F^{-1}-\theta)}{\mu_M-(1-r)\rho\theta} \left(e^{-(1-r)\rho\theta (m\tau+s)}-e^{-\mu_M(m\tau+s)}\right) \Big). 
\end{split}\end{equation*}
By multiplying by $e^{\mu_S(m\tau+s)}$  both sides of latter inequality, we get
\begin{equation*}
\begin{split}
&\Lambda_{np+m} \tau e^{\mu_S m\tau } + \Lambda_{np+m-1} \tau e^{\mu_S (m-1)\tau} + \dots + \Lambda_{np} \tau  + M_S(np\tau)
\\
&\geq \kappa_{\max}^{np+m}\Big( M(np\tau) e^{(\mu_S-\mu_M)(m\tau+s)} \\
&+ F(np\tau) \frac{r\rho(\mathcal{N}_F^{-1}-\theta)}{\mu_M-(1-r)\rho\theta} \left(e^{(\mu_S-(1-r)\rho\theta) (m\tau+s)}-e^{(\mu_S-\mu_M)(m\tau+s)}\right) \Big).
\end{split}
\end{equation*}
This inequality gives the strongest condition when $s=\tau$. Thus, we enforce,
\begin{equation*}
\begin{split}
&\Lambda_{np+m} \tau e^{\mu_S m\tau } + \Lambda_{np+m-1} \tau e^{\mu_S (m-1)\tau} + \dots + \Lambda_{np} \tau  + M_S(np\tau)
\\
&\geq \kappa_{\max}^{np+m}\Big( M(np\tau) e^{(\mu_S-\mu_M)(m+1)\tau} \\
&+ F(np\tau) \frac{r\rho(\mathcal{N}_F^{-1}-\theta)}{\mu_M-(1-r)\rho\theta} \left(e^{(\mu_S-(1-r)\rho\theta) (m+1)\tau}-e^{(\mu_S-\mu_M)m\tau}\right) \Big).
\end{split}
\end{equation*}
We get, for $m=0,\dots,p-1$,
\begin{equation*}
\begin{split}
&\Lambda_{np+m} \tau \geq e^{-\mu_S m\tau }\Big[ - \Lambda_{np+m-1} \tau e^{\mu_S (m-1)\tau} - \dots - \Lambda_{np} \tau  - M_S(np\tau)
\\
&+ \kappa_{\max}^{np+m}\Big( M(np\tau) e^{(\mu_S-\mu_M)(m+1)\tau} \\
&+ F(np\tau) \frac{r\rho(\mathcal{N}_F^{-1}-\theta)}{\mu_M-(1-r)\rho\theta} \left(e^{(\mu_S-(1-r)\rho\theta) (m+1)\tau)}-e^{(\mu_S-\mu_M)m\tau}\right) \Big) \Big].
\end{split}
\end{equation*}
The result below follows.

\begin{thm}[Stabilization by impulsive control with sparse measurements] 
\label{th2} 
Let $p\in\mathbb{N}^{*}$,  $\epsilon \geq 0$ and $\theta>0$ such that $\theta+\epsilon < \min\left(1,\dfrac{\mu_S}{\mu_F}\right)\mathcal{N}_{F}^{-1}$. Assume that, for any $n\in\mathbb{N}$, $m=0,1,\dots,p-1$,
\[
\tau\Lambda_{np+m}\geq-M_{S}(np\tau)e^{-m\mu_{S}\tau}-\tau\sum_{i=0}^{m-1}\Lambda_{np+i}e^{-\left(m-i\right)\mu_{S} \tau}+K_{p,\epsilon}\left(\begin{array}{c}
M(np\tau)\\
F(np\tau)
\end{array}\right).
\]
with
\[
K_{p,\epsilon}
\coloneqq\dfrac{e^{\mu_{S}\tau}}{\gamma}
\dfrac{e^{-\beta(M_{\min}^{np+m}+F_{\min}^{np+m})}-\left(\mathcal{N}_{F}^{-1}-\theta\right)}{\left(\mathcal{N}_{F}^{-1}-\theta\right)-\epsilon e^{-\beta(M_{\min}^{np+m}+F_{\min}^{np+m})}}
V_{p,\epsilon}^T,
\]
where $M_{\min}^{np+m}$ and $F_{\min}^{np+m}$ were introduced in \eqref{MFmin} and
\[
V_{p,\epsilon}\coloneqq \left(\begin{array}{c}
e^{-\mu_{M}m\tau}M(np\tau)\\
\dfrac{r\rho\left(\mathcal{N}_{F}^{-1}-\theta\right)}{\mu_{M}-(1-r)\rho\theta}\left(e^{-(1-r)\rho\theta m\tau}-e^{-\mu_{M}m\tau}\right)F(np\tau)
\end{array}\right)
\]
Then, every solution of system \eqref{eq:sit_cste} converges exponentially towards $\mathbf{0}$,
with a convergence speed bounded from below by a value independent
of the initial condition.

If moreover 
\[
\tau\Lambda_{np+m}\leq K_{p,\epsilon}\begin{pmatrix}M(np\tau)\\
F(np\tau)\ensuremath{}
\end{pmatrix}
\]
then the series of impulses $\sum_{n=0}^{+\infty}\Lambda_{n}$ converges.

\end{thm}
\subsection{Mixed impulsive control strategies}
As done in \cite{Bliman2019}, we can combine the open-loop and the closed-loop controls in order to derive the best strategy that will use fewer sterile males. More precisely, at each release time, we compute the open and closed-loop controls, and we choose the smaller one. See \cite[Section 6]{Bliman2019} for more details.

\section{Numerical simulations}
We present several numerical simulations to illustrate our results. In particular, we  compare the linear and the nonlinear feedback control laws, as well as mixed control strategies.

\subsection{{\em Aedes Albopictus} parameters}
Parameters estimate is based on several publications \cite{Damiens2016,Dumont2012,Oliva,LeGoff2019}. In particular, we estimate the characteristic $\beta$ of the competition effect taking into account the population estimates obtained in \cite{LeGoff2019}: around  $6,000$ males during the rainy season and, $600$ males during the dry season.  
\begin{table}[h!]
\centering
\begin{tabular}{|c|c|l|}   \hline
\textbf{Par.} & \textbf{Value}	 & \textbf{Description} \\ \hline
$\rho$		    	&0.9*0.74*10=6.66& \shortstack{Number of viable eggs (that reach the adult stage)\\
a female can deposit per day} \\ \hline
$r$							&0.5& \shortstack{$r:(1-r)$ expresses the primary sex ratio \\ among offsprings}  \\ \hline
$\sigma$	    	&0.05&	\shortstack{Regulates the larvae development into adults under \\  density dependence and larval competition} \\ \hline
$K$	          	&165.21 &	Carrying capacity in the rainy season\\ \hline
$\mu_M $	    	&1/13&	Mean mortality rate of wild adult male mosquitoes \\ \hline
$\mu_F $	    	&1/15&	Mean mortality rate of wild adult female mosquitoes \\ \hline
$\mu_S$				&1/8.5&	Mean mortality rate of sterile adult male mosquitoes \\ \hline 
$\ensuremath{\gamma}$ &0.91 & Competitiveness index of sterile male mosquitoes \cite{lyaloo2019}\\ \hline 
\end{tabular}
\caption{{\em Aedes albopictus} parameters values (estimated from \cite{Damiens2016,Delatte2009,Dumont2012,Oliva,LeGoff2019,lyaloo2019})}
\label{tab1}
\end{table}
According to Table \ref{tab1},  we derive $\cN_F\approx 49.95$ and $\cN_M\approx 43.29$. The basic offspring number $\cN_F$ is pretty large but realistic in tropical context. According to our previous result, we need to impose $\epsilon< \mathcal{N}_F^{-1},$ then individual fertility in the sterile male population has to be lower than $2 \%$. If not, if for instance $\epsilon\approx 5\%$ then, according to \eqref{lower_bound}, $M^*_\ell=1,404,F^*_\ell=1,620$ individuals: this value is reached only for very large releases value, i.e. $\Lambda>10^{10}$, that are completely unrealistic. Altogether, even with very massive releases, the population reduction is only of $76.5 \%$ which is not sufficient to reduce the epidemiological risk.

Then, assuming $M^*\approx 6,000$ individuals (in parenthesis we write the corresponding amount for the dry season), the global competition coefficient $\beta= \dfrac{\sigma}{K}=3.482 \times 10^{-4}$. At equilibrium, $E^*=(M^*,F^*)$, the mosquito population verifies $M^*= 6,000$ and $F^*\approx 6,923$ individuals per hectare. 

When $\epsilon=0$, for open-loop periodic impulsive releases carried out every $7$ (resp.\ $14$) days, we consider the release value given in (\ref{eq:cond}), page \pageref{eq:cond}, to estimate the minimum of sterile males to release, that is, $ \tau \left(\lfloor \Lambda_{crit}^{per}\rfloor+1\right)=7\times 8,304=58,128$ (resp.\ $14\times 9,792=137,088$) sterile males per hectare and per week (resp.\ every two weeks). Note also that for the weekly (every $14$ days) release, we approximately release $10$ ($23$) times more sterile males than wild males. In fact, we recover the (minimal) amount of sterile males that is usually recommended by the International Atomic Energy Agency (IAEA).

When $\epsilon=0.015>0$, the open-loop control requires to release at least $\tau \left(\lfloor \Lambda_{crit}^{per}\rfloor+1\right)=7\times 32,619=228,333$ (resp.\ $14\times 38,469=538,566$) sterile males per hectare and per week (resp.\ every two weeks). It is interesting to notice the rise in the release size even with a small residual fertility: we need to release almost $4$ times more sterile males. Thus, it is preferable to reduce or avoid the residual fertility all along the experiment if we consider only open-loop control. We will study later the impact on mixed-control strategies.

\subsection{{\em Bactrocera dorsalis} parameters}
To estimate the parameters we rely on several publications, like  \cite{Ekesi2006,Shelly2010,Salum2014,Pieterse2019,Yusof2019}. However, {\em Bactrocera dorsalis} has a rapid dynamics depending on the type of fruits it develops, such that its basic offspring number can vary from $100$ to $500$ \cite{Ekesi2006, Salum2014,Pieterse2019}. 
From Table \ref{tab2}, we get that $\cN_M\approx 251.42$ and $\cN_F\approx 232.06$.

Population estimate for \textit{Bactrocera dorsalis} are much more difficult to find in the literature than for mosquitoes. However, in \cite{Tan1994} the male population was estimated between $3,300$ and $18,000$.. Like for mosquitoes, seasonal variation can also occur.
Thus, assuming the male population around $6,000$ individuals per hectare, we can deduce $\beta$ and then, setting $\sigma=0.05$, estimate $K$. We get $\beta \approx 4.7210\times 10^{-4}$ and $K=106$.

\begin{table}[h!]
\centering
\begin{tabular}{|c|c|l|}   \hline
\textbf{Parameter} & \textbf{Value}	 & \textbf{Description} \\ \hline
$\rho$		    	&6.0& 
\shortstack{Number of viable eggs (that reach the adult stage)\\
a female can deposit per day} \\ \hline
$r$							&0.485& $r:(1-r)$ expresses the primary sex ratio in offspring  \\ \hline
$\sigma$	    	&0.05&	
\shortstack{Regulates the larvae development into adults under \\ density dependence and larval competition} \\
\hline
$K$	          	&106&	Carrying capacity \\ \hline
$\mu_M $	    	&1/86.4&	Mean mortality rate of wild adult male fruit flies \\ \hline
$\mu_F $	    	&1/75.1&	Mean mortality rate of wild adult female fruit flies \\ \hline
$\mu_S$				&1/86.4&	Mean mortality rate of sterile adult fruit flies \\ \hline 
$\ensuremath{\gamma}$ &0.6 & Competitiveness index of sterile male fruit flies \\ \hline 
\end{tabular}
\caption{{\em Bactrocera dorsalis} parameters values (estimated from \cite{Ekesi2006}) on Mango; see also \cite{Yusof2019} for the SIT parameters}
\label{tab2}
\end{table}

The minimal Gamma irradiation dose such that the lifespan of irradiated/treated flies is almost similar to untreated flies is $100$ Gy \cite{Yusof2019}. Also, the 100 Gy treatment seems to be sufficient to induce $100\%$ sterility (see   \cite[Table 2 page 4]{Yusof2019}).

Despite the fact that the lifespan of the sterile males is large, weekly massive releases are recommended in real experiments. This is mainly due to the fact that the dynamics of \textit{B. dorsalis} is strong. Thus, for a weekly open-loop release strategy, the number of sterile males to release could be, for instance, $2 \times \tau \left(\lfloor \Lambda_{crit}^{per}\rfloor+1\right)=2\times 7\times 3,494=48,916$. Compared to the mosquito case, and since the basic offspring number is very large, the critical value seems to be low. This is thank to the lifespan of the sterile male being large, regardless  of the bad competitive index.

\subsection{Simulations with full sterility, i.e. $\epsilon=0$}
We apply the long term control strategy (introduced in Subsection \ref{SSbox}) which consists in setting a desired long term release size $M_{S,{\rm obj}}$, then computing the corresponding value of the threshold $\bf{E}_1$ and performing releases in two stages. A first stage with massive releases (either open or closed-loop control, or a combination of both) in order to enter the box $[\bf{0},\bf{E}_1[$ and a second {\em long term} stage of releases of constant size $M_{S,{\rm obj}}$.

We first start with $100\%$ sterility and compare the results obtained with linear and nonlinear feedback controls. We consider only periodic releases with two different periods: $\tau=7$ and $\tau=14$, and we assume to get estimates of the wild population every $p\tau$ days, for $p=1$ or $p=4$. We can consider several choices for $\theta>0$ as long as $\theta+\epsilon< \min\left(1,\frac{\mu_S}{\mu_F}\right)\cN_F^{-1}$, where we take $\epsilon=0$ in this subsection. We consider 2 values for $\theta$: ${0.99}\cN_F^{-1}$ and ${0.2}\cN_F^{-1}$. We now provide the time needed to enter the box $[\bf{0},\bf{E}_1[$, for each period $\tau$, each $p$ and each choice of $\theta$. 
While the case $\theta={0.99}\cN_F^{-1}$ guarantees a faster convergence of $F$ to zero, this does not necessarily imply the best outcome in terms of released insects necessary to enter the box $[\bf{0},\bf{E}_1[$.
\subsubsection{The \textit{Aedes albopictus} case}
We choose $M_{S,{\rm obj}}=100$, leading to ${\bf{E}_1}=(1.45,1.67)$ when $\tau=7$, and ${\bf{E}_1}=(0.44,0.51)$ when $\tau=14$. We could choose a larger value for $M_{S,{\rm obj}}$, but this is to show the release ``effort" that is necessary event to control a very small population.

\begin{figure}[!h]
\begin{tabular}{cc}
 \includegraphics[width=.5\textwidth]{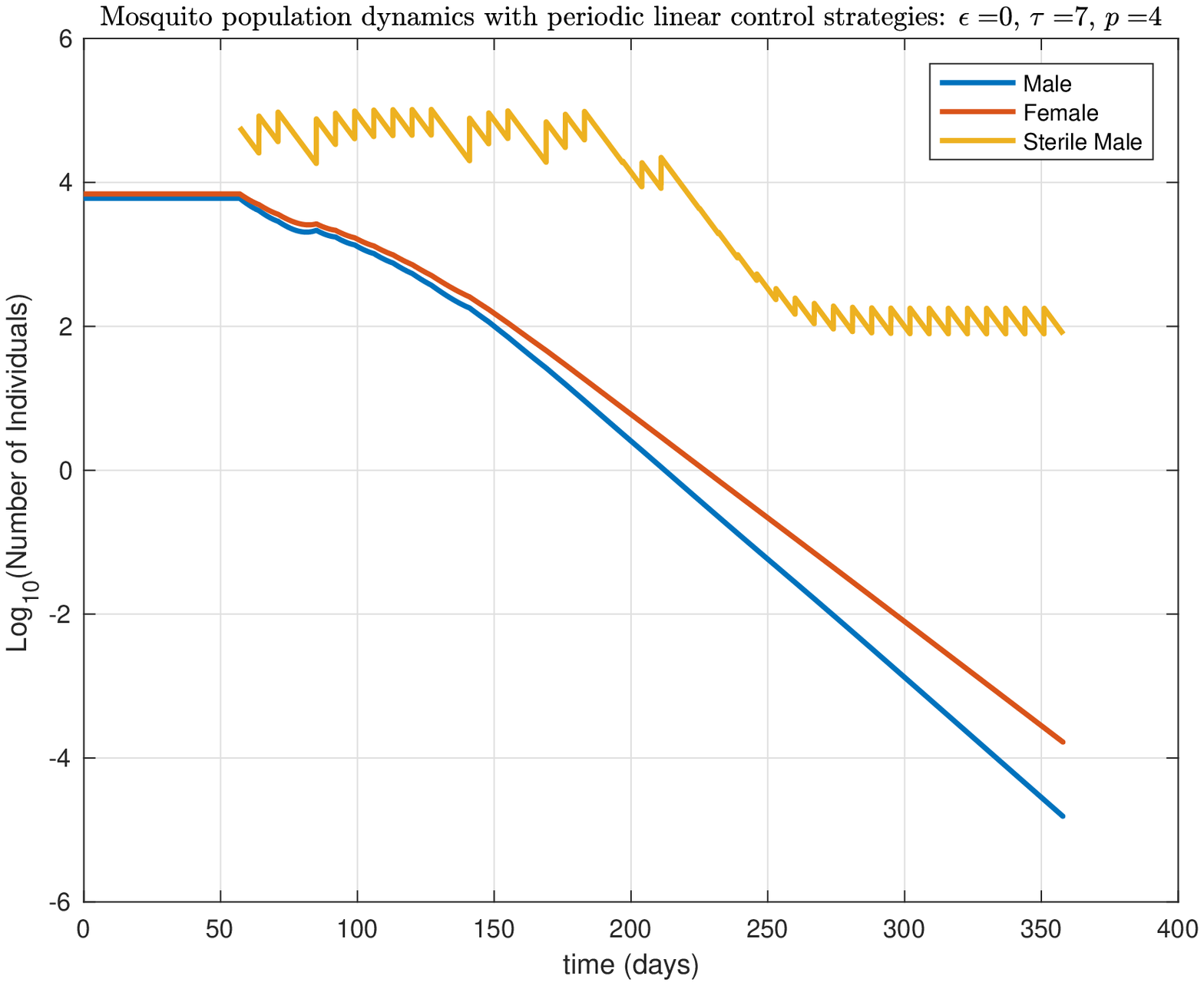} &
 \includegraphics[width=.5\textwidth]{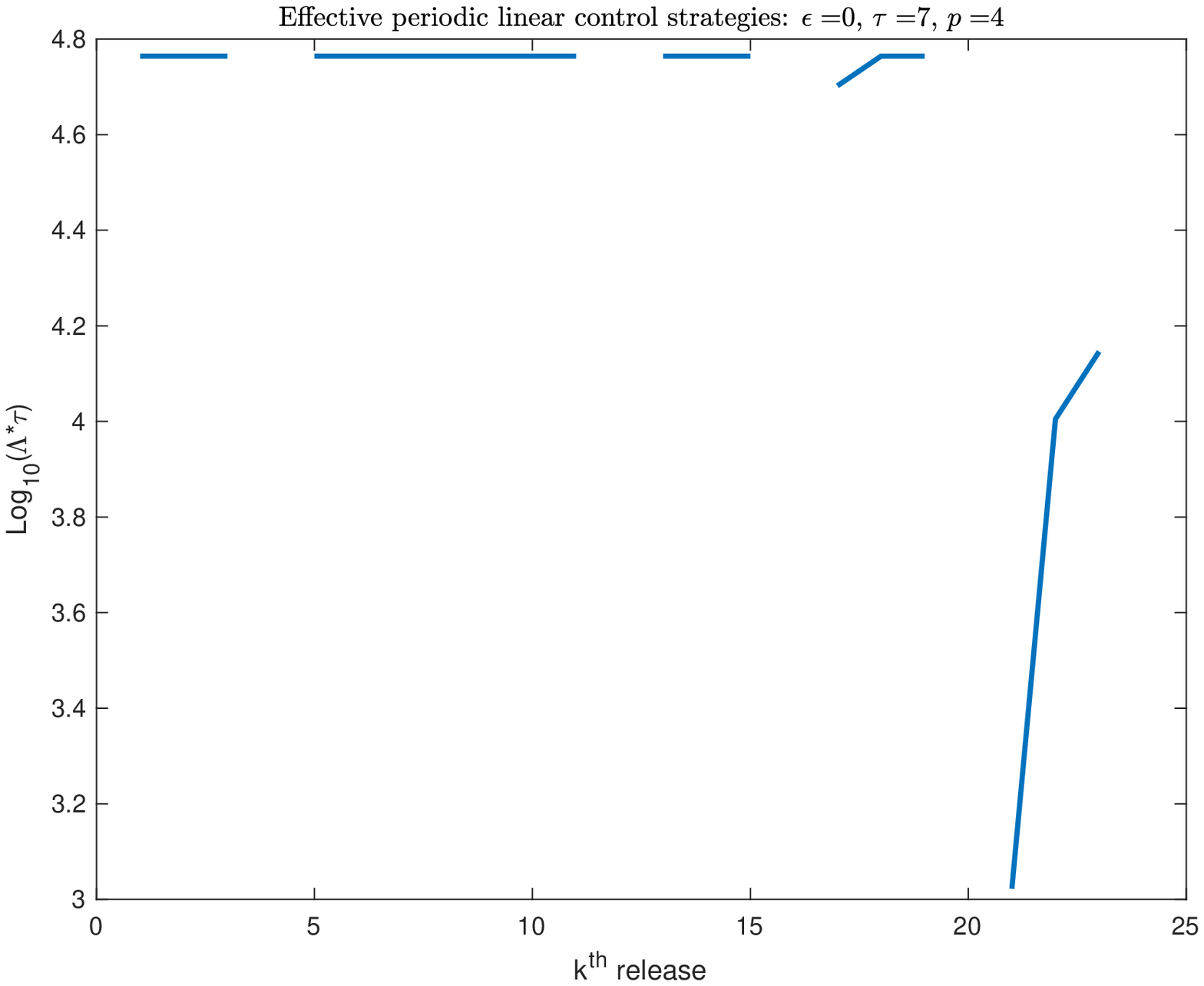} \\
 (a) & (b) \\
  \includegraphics[width=.5\textwidth]{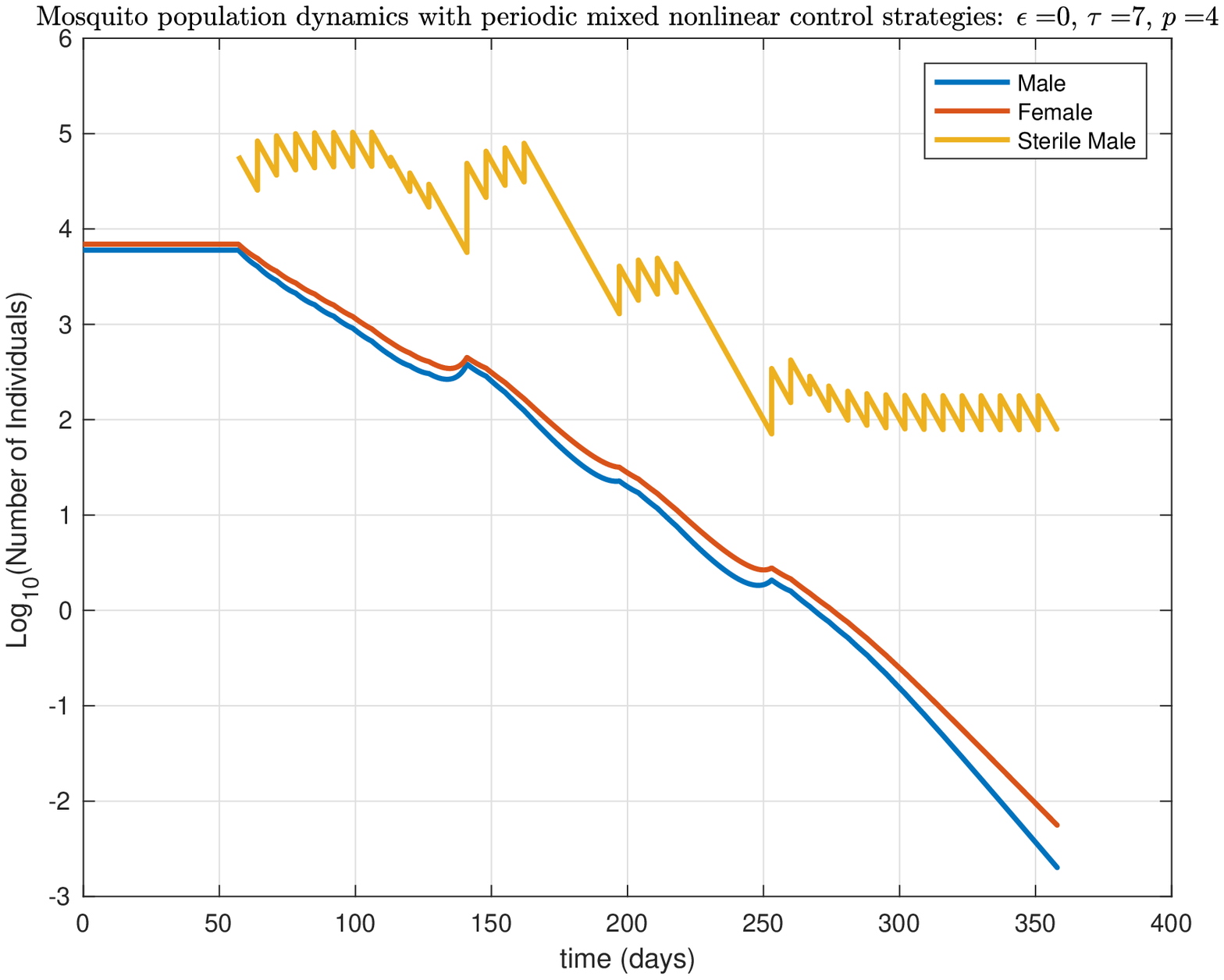} &
 \includegraphics[width=.5\textwidth]{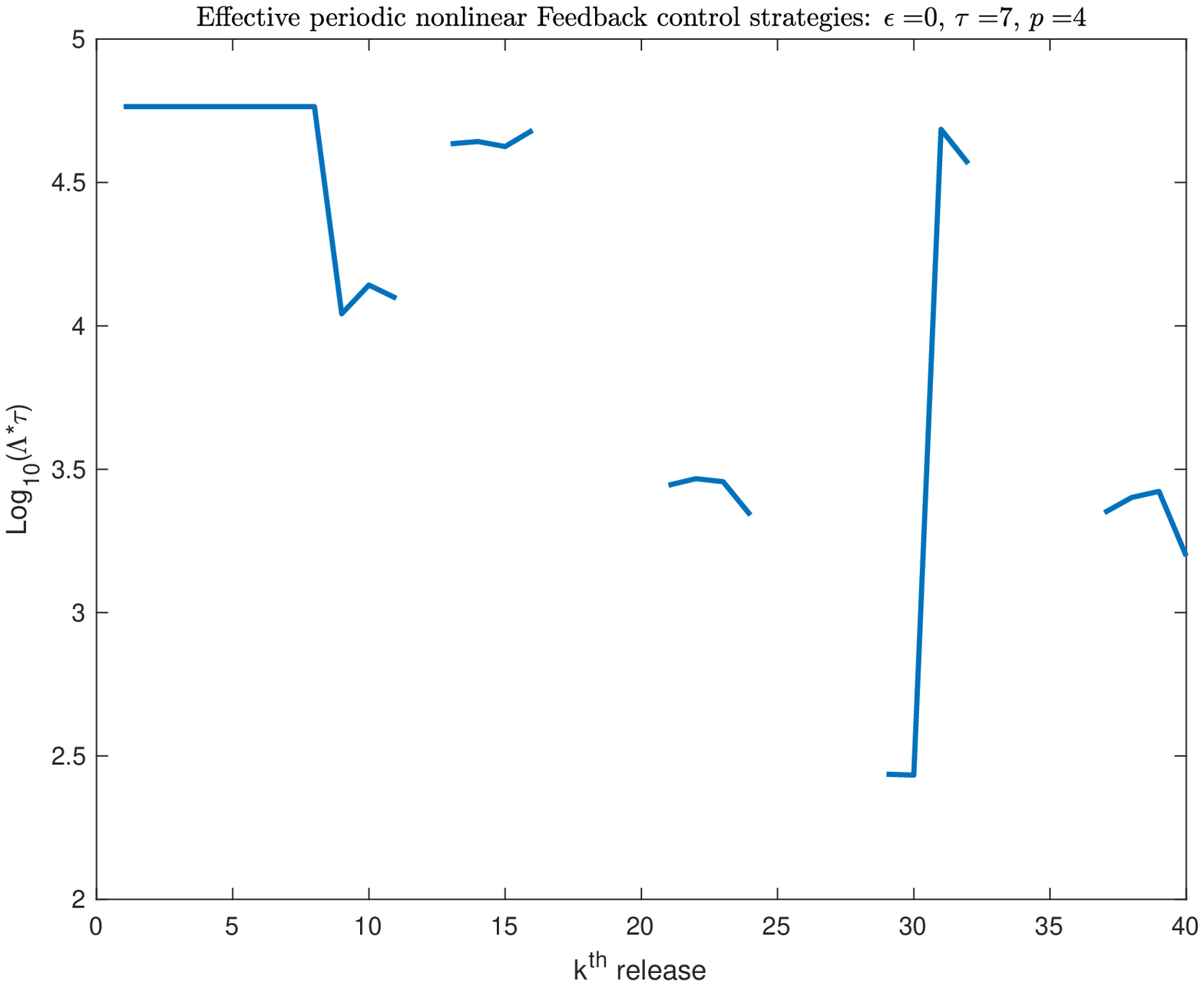} \\
 (c) & (d)
 \end{tabular}
 \caption{\textit{Aedes albopictus} - Mixed periodic impulsive  SIT control of system (\ref{eq:periodicTIS}) with $\theta=0.2 {\cN_F^{-1}}$, $\tau=7$ and $p=4$ - {\bf Linear} control: (a) population dynamics; (b) Field releases timing - {\bf Nonlinear} control: (c) population dynamics; (d) Field releases timing. See Tables \ref{tab2a} and \ref{tab2b}, page \pageref{tab2a}.}
 \label{fig:1}
\end{figure}

In Tables \ref{tab1a}, \ref{tab1b}, \ref{tab2a} and  \ref{tab2b} we show some results for a $400$-day mixed control. Clearly, the choice of $\theta$ has a direct influence on the cumulative number of sterile males and the number of massive releases. When $p>1$, the best results are obtained with the nonlinear mixed control, simply because the first releases are smaller (compare (a) and (b) or (c) and (d) in Fig. \ref{fig:1}, page \pageref{fig:1}). Overall, and taking into account that sparse measurements occur every $4$ weeks,  the best strategy is the $7$-days release strategy seems to be the most appropriate: the lowest number of insects to release combined with only $21$ ``massive" mixed releases (see Fig. \ref{fig:1}) to reach the box $[\bf{0},\bf{E}_1[$. Note that the nonlinear mixed control needs $33\%$ less sterile males than the linear mixed control.

As expected by Remark \ref{rem2}, page \pageref{rem2}, the parameter $\theta$ has an impact on the duration of the SIT treatment. However, when $p=4$ and $\theta={0.2}{\cN_F^{-1}}$, the duration is almost the same for both the linear and nonlinear mixed controls, with a certain gain on the number of sterile males to release with nonlinear mixed control.

Fig. \ref{fig:1} provides a typical output of a mixed-control strategy. We show the results for the linear and nonlinear mixed controls (in open and closed-loop): the difference between both approaches occurs in the beginning, where in the linear control large amount of sterile insects (open-loop control) are released. Figs. \ref{fig:1} (b) and (d) show the times of  releases: as seen, the releases do not occur every $\tau$ days, but only if the size of the sterile males is not sufficient to continue to drive the wild population to extinction. This may depend on the periodicity of the releases but also on the vital parameters related to the sterile males. However, with a high mortality rate, $\mu_S=1/8.5$, even $14$-days periodic releases can work, but this require to release a larger number of sterile males.

\begin{table}[!th]
\centering
\begin{tabular}{|c|c|c|c|}   \hline
$p$ & Period (days) & Cumulative number of & Number of effective  \\
&  & released sterile males &  releases to reach $[\bf{0},\bf{E}_1[$ \\
 \hline
$1$ &$\tau=7$  & $  9.87560 \times 10^5$ & $22$\\
\hline
& $\tau=14$  &  $1.181635\times 10^6$  & $12$\\
\hline
 \hline
$4$&$\tau=7$  & $ 1.162560 \times 10^6$ & $20$\\
\hline
& $\tau=14$  &  $1.423067 \times 10^6$  & $11$\\
\hline
\end{tabular}
\caption{\textit{Aedes albopictus} - Cumulative number of released sterile males and number of releases for \textbf{linear} mixed control, when $\theta=0.99{\cN_F^{-1}}$}
\label{tab1a}
\end{table}
\begin{table}[!th]
\centering
\begin{tabular}{|c|c|c|c|}   \hline
$p$ & Period (days) & Cumulative Number of  & Nb of effective releases \\
&  & released sterile males & to reach $[\bf{0},\bf{E}_1[$ \\
 \hline
$1$ &$\tau=7$  & $9.87164 \times 10^5$ & $22$\\
\hline
& $\tau=14$  &  $1.181335 \times 10^6$  & $12$\\
\hline
 \hline
$4$&$\tau=7$  & $ 1.001690 \times 10^6$ & $18$\\
\hline
& $\tau=14$  &  $1.297330 \times 10^6$  & $11$\\
\hline
\end{tabular}
\caption{\textit{Aedes albopictus} - Cumulative number of released sterile males and number of releases and number of releases for \textbf{nonlinear} mixed control, when $\theta=0.99 {\cN_F^{-1}}$}
\label{tab1b}
\end{table}
\begin{table}[!th]
\centering
\begin{tabular}{|c|c|c|c|}   \hline
$p$ & Period (days) & Cumulative Number of  & Nb of effective releases \\
&  & released sterile males & to reach $[\bf{0},\bf{E}_1[$ \\
 \hline
$1$&$\tau=7$  & $ 5.81934 \times 10^5$ & $33$\\
\hline
& $\tau=14$  &  $7.41805 \times 10^5$  & $15$\\
\hline
 \hline
$4$&$\tau=7$  & $ 9.47466 \times 10^5$ & $19$\\
\hline
& $\tau=14$  &  $1.412932 \times 10^6$  & $13$\\
\hline
\end{tabular}
\caption{\textit{Aedes albopictus} - Cumulative number of released sterile males and number of releases for \textbf{linear} mixed  control, when $\theta=0.2 {\cN_F^{-1}}$}
\label{tab2a}
\end{table}
\begin{table}[!th]
\centering
\begin{tabular}{|c|c|c|c|}   \hline
$p$ & Period (days) & Cumulative Number of  & Nb of effective releases \\
&  & released sterile males & to reach $[\bf{0},\bf{E}_1[$ \\
 \hline
$1$&$\tau=7$  & $ 5.71501 \times 10^5$ & $34$\\
\hline
& $\tau=14$  &  $7.28663 \times 10^5$  & $15$\\
\hline
 \hline
$4$&$\tau=7$  & $ 6.91024 \times 10^5$ & $21$\\
\hline
& $\tau=14$  &  $8.98257 \times 10^5$  & $10$\\
\hline
\end{tabular}
\caption{\textit{Aedes albopictus} - Cumulative number of released sterile males and number of releases for \textbf{nonlinear} mixed control, when $\theta=0.2 {\cN_F^{-1}}$}
\label{tab2b}
\end{table}

As done in \cite{Yatat2020}, we consider that another control methods should be used, like for instance one week of adulticide can be implemented before SIT starts, as recommended by the IAEA. Thus, taking $\tau=7$ and $p=4$, we obtain the outcome given in Table \ref{tab_adulticide}.
\begin{table}[!th]
\centering
\begin{tabular}{|c|c|c|c|c|}   \hline
Mixed-Control & $p$ & Period (days) & Cumulative Number of  & Nb of effective Releases \\
& &  & released sterile males &  to reach  $[\bf{0},\bf{E}_1[$ \\
 \hline
Linear & $4$ & $\tau=7$  &  $6.07062 \times 10^5$  & $13$\\
 \hline
Nonlinear  & $4$ & $\tau=7$  &  $3.12157 \times 10^5$  & $11$\\
\hline
\end{tabular}
\caption{\textit{Aedes albopictus} - Cumulative number of released sterile males and number of releases for linear or {\bf nonlinear} mixed control, after one week of adulticide, when $\theta=0.2 \cN_F^{-1}$}
\label{tab_adulticide}
\end{table}

Clearly, comparing Tables \ref{tab_adulticide} and \ref{tab2b}, the gain is substantial in terms of sterile males to release (almost $55\%$ less) and also in terms of effective releases ($10$ less).

\subsubsection{The \textit{Bactrocera dorsalis} case}
We choose $M_{S,{\rm obj}}=2000$, leading to ${\bf{E}_1}\approx (65.31, 60.28)$ when $\tau=7$ days, and ${\bf{E}_1}\approx (30.35, 28.01)$ when $\tau=14$ days.
We consider $\theta={0.3}{\cN_F^{-1}}<\min\left(1,\frac{\mu_S}{\mu_F}\right){\cN_F^{-1}}$, $p=4$ or $p=8$.
For the open-loop control we assume that we release $2\times \tau \times \Lambda^{per}_{\rm crit}$ sterile males.

\begin{table}[!th]
\centering
\begin{tabular}{|c|c|c|c|}   \hline
$p$ & Period (days) & Cumulative Number of  & Nb of effective releases \\
&  & released sterile males & to reach $[\bf{0},\bf{E}_1[$ \\
 \hline
$4$ &$\tau=7$  & $ 2.204530 \times 10^6$ & $59$\\
\hline
& $\tau=14$  &  $2.123689\times 10^6$  & $30$\\
\hline
 \hline
$8$&$\tau=7$  & $2.332524\times 10^6$  & $47$\\
\hline
& $\tau=14$  &  $2.434298 \times 10^6$  & $25$\\
\hline
\end{tabular}
\caption{\textit{Bactrocera dorsalis} - Cumulative number of released sterile males and number of releases for \textbf{linear} mixed control, when $\theta=0.3/ \cN$}
\label{tabBD1}
\end{table}

\begin{table}[!th]
\centering
\begin{tabular}{|c|c|c|c|}   \hline
$p$ & Period (days) & Cumulative Number of  & Nb of effective releases \\
&  & released sterile males & to reach $[\bf{0},\bf{E}_1[$ \\
 \hline
$4$ &$\tau=7$  & $ 1.891870 \times 10^6$ & $63$\\
\hline
& $\tau=14$  &  $1.816422 \times 10^6$  & $35$\\
\hline
 \hline
$8$&$\tau=7$  & $1.969639 \times 10^6$  & $48$\\
\hline
& $\tau=14$  &  $2.388379 \times 10^6$  & $25$\\
\hline
\end{tabular}
\caption{\textit{Bactrocera dorsalis} - Cumulative number of released sterile males and number of releases for {\bf nonlinear} mixed control, when $\theta=0.3 \cN_F^{-1}$}
\label{tabBD2}
\end{table}

The results, given in Tables \ref{tabBD1} and \ref{tabBD2}, are not surprising: in almost all releases, open-loop releases occur. However, in the case $\tau=14$ and $p=4$, then the mixed-nonlinear control is interesting, with a gain of almost $15\%$ thanks to the mixed-linear control. For a large area, this request to be able to manufacture billions of sterile males.

These results confirm that control by SIT alone is almost impossible for \textit{Bactrocera dorsalis}. Additional control tools (like female trapping or combination with Methyl-Eugenol \cite{Shelly2010}) are necessary. Here, as in \cite{Yatat2020}, we consider a one week treatment with $100\%$ efficiency.

\begin{table}[!th]
\centering
\begin{tabular}{|c|c|c|c|c|}   \hline
Mixed-Control & $p$ & Period (days) & Cumulative Number of  & Nb of effective releases \\
& &  & released sterile males & to reach $[\bf{0},{\bf{E}_1}[$ \\
 \hline
Linear & $4$ & $\tau=14$  &  $1.448964 \times 10^6$  & $23$\\
 \hline
Nonlinear  & $4$ & $\tau=14$  &  $1.661831 \times 10^6$  & $31$\\
\hline
\end{tabular}
\caption{\textit{Bactrocera dorsalis} - Cumulative number of released sterile males and number of releases for \textbf{linear} or \textbf{nonlinear} mixed control, after one week of adulticide}
\label{tabBD3}
\end{table}
According to Table \ref{tabBD3}, there is a clear improvement in the gain of the number of releases and thus in total number of the released sterile males. Also, linear control is better, but in fact this depend on the choice of the open-loop control release. 

Since $\cN_F$ is very large for \textit{Bactrocera dorsalis}, the residual fertility should be not greater than $0.0043$. Assuming that $\epsilon=0.01$ (i.e. only $1\%$ of residual fertility) then, according to our previous estimate, whatever the size of the releases, the male population can not go down under $\dfrac{\ln(\epsilon \cN_F)}{c}$. According to the parameters values, this leads to a minimal population of $928$ males per ha and, using relation (\ref{MF}), to a minimal population of $856$ females per ha. These lower bounds will not be reached even for very large but still realistic releases.

\subsection{The Non-fully sterile case}
We consider only the partial-sterile case for \textit{Aedes albopictus}. We assume a residual fertility of $1.5\%$, i.e. $\epsilon=0.015$. We choose $\theta=0.2/\cN_F$ such that $\theta+\epsilon<1/\cN_F$.

In that case, for  open-loop periodic impulsive releases carried out every 7 (resp.  14) days, we estimate that we need to release $7 \times 32,619 =228,333$ ($14 \times 38,469=538,566$) sterile males per ha and per week. Compare to the $\epsilon=0$ case, the amount of insects to release is $4$ times larger.

\begin{table}[!th]
\centering
\begin{tabular}{|c|c|c|c|}   \hline
$p$ & Period (days) & Cumulative Number of  & Nb of effective releases \\
&  & released sterile males &  to reach $[\bf{0},\bf{E}_1[$ \\
 \hline
$1$ &$\tau=7$  & $ 5.258893 \times 10^6$ & $74$\\
\hline
& $\tau=14$  &  $6.870483\times 10^6$  & $42$\\
\hline
 \hline
$4$&$\tau=7$  & $ 1.2146326 \times 10^7$ & $54$\\
\hline
& $\tau=14$  &  $1.5181271 \times 10^7$  & $32$\\
\hline
\end{tabular}
\caption{Residual fertility case - \textit{Aedes albopictus} - Cumulative number of released sterile males and number of releases for \textbf{linear} closed-loop control, when $\theta=0.2\cN_F^{-1}$ and $\epsilon=0.015$.}
\label{tabPS1a}
\end{table}
\begin{table}[!th]
\centering
\begin{tabular}{|c|c|c|c|}   \hline
$p$ & Period (days) & Cumulative Number of  & Nb of effective releases \\
&  & released sterile males &  to reach $[\bf{0},\bf{E}_1[$ \\
 \hline
$1$&$\tau=7$  & $ 3.443861 \times 10^6$ & $78$\\
\hline
& $\tau=14$  &  $6.151367\times 10^6$  & $42$\\
\hline
 \hline
$4$&$\tau=7$  & $ 4.48830 \times 10^6$ & $58$\\
\hline
& $\tau=14$  &  $6.549222 \times 10^6$  & $42$\\
\hline
\end{tabular}
\caption{Cumulative number of released sterile males and number of releases for \textbf{nonlinear} closed-loop control, when $\theta=0.2 \cN_F^{-1}$ and $\epsilon=0.015$.}
\label{tabPS1b}
\end{table}
\begin{figure}[!t]
\begin{tabular}{cc}
 \includegraphics[width=.5\textwidth]{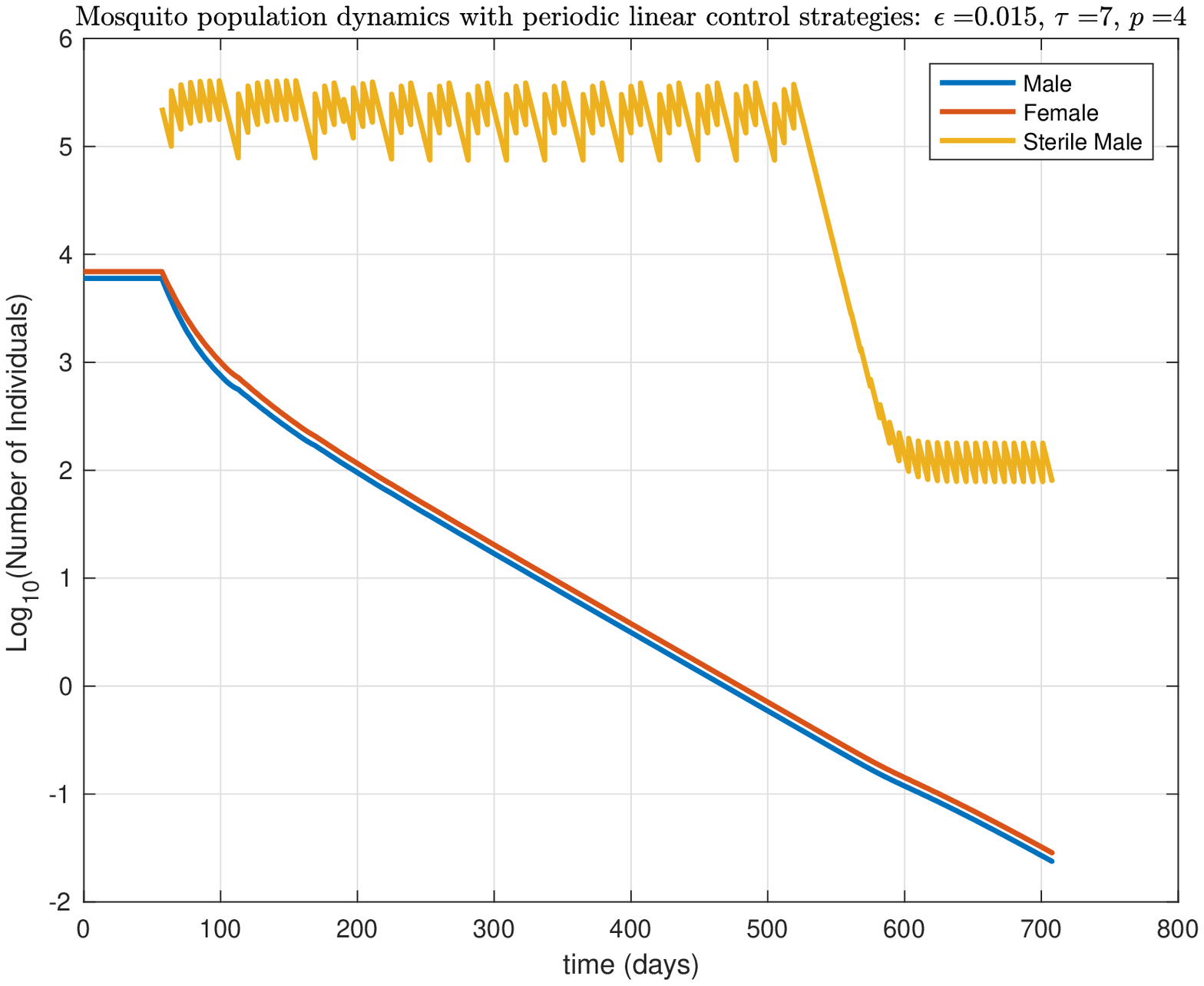} &
 \includegraphics[width=.5\textwidth]{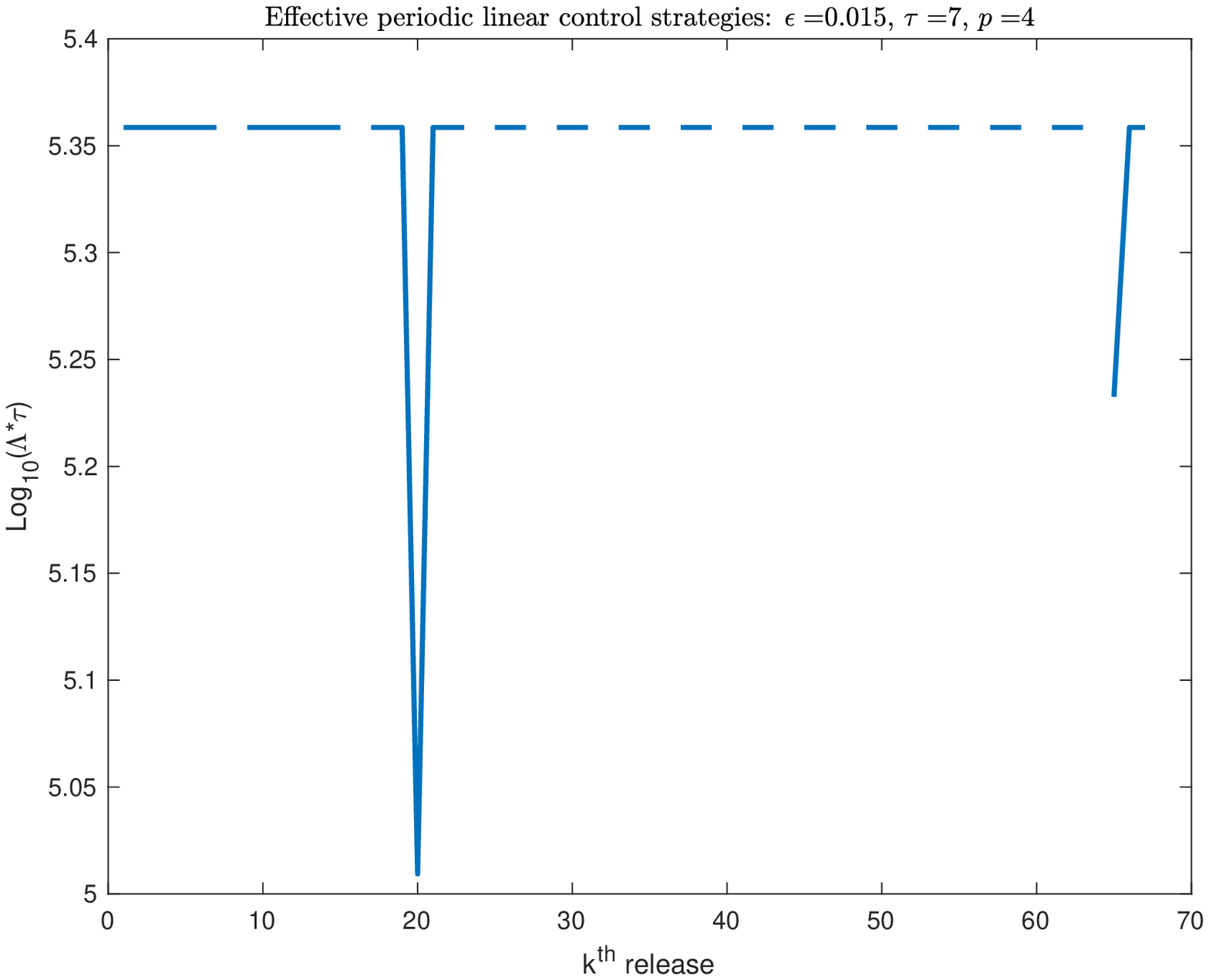} \\
 (a) & (b) \\
  \includegraphics[width=.5\textwidth]{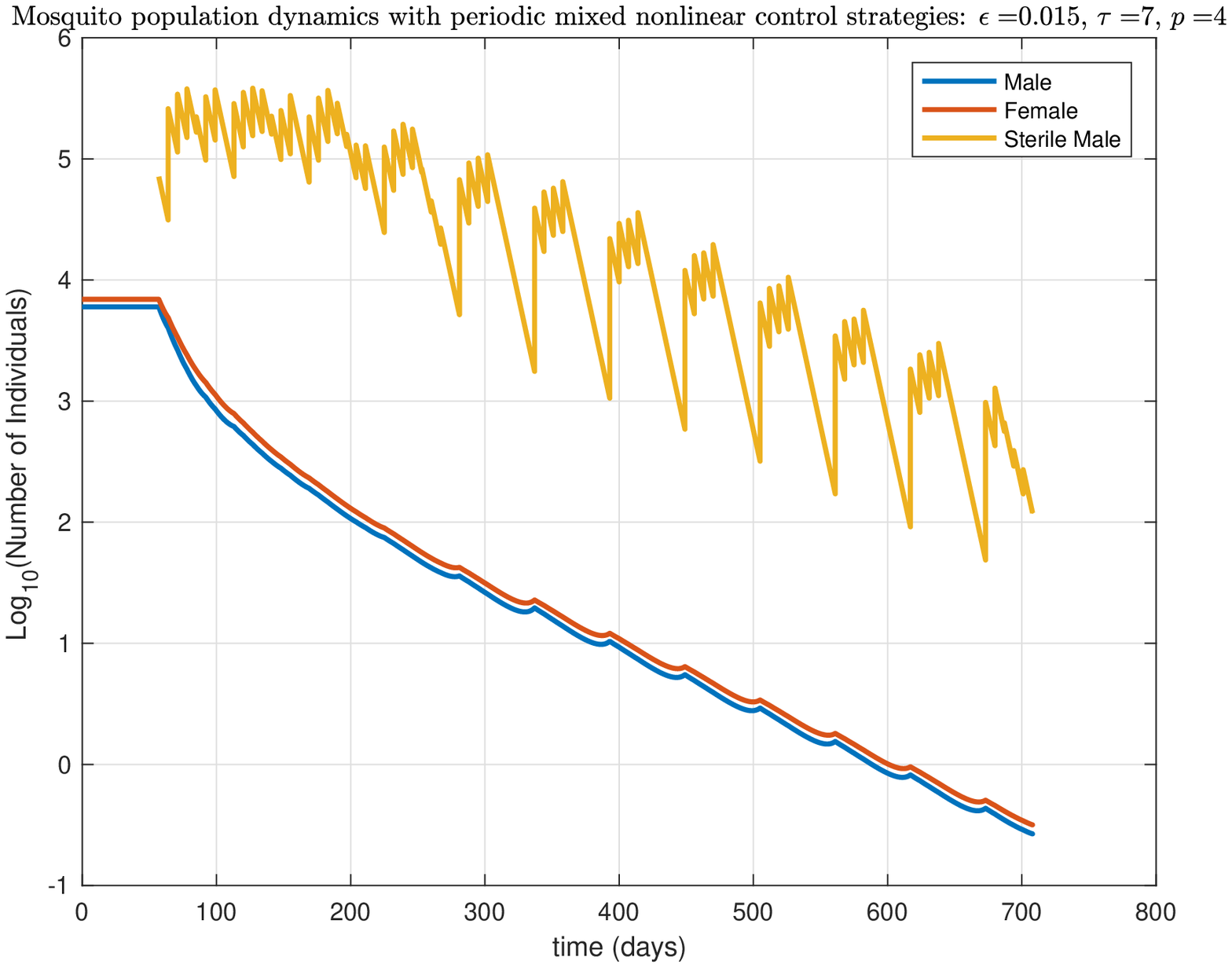} &
 \includegraphics[width=.5\textwidth]{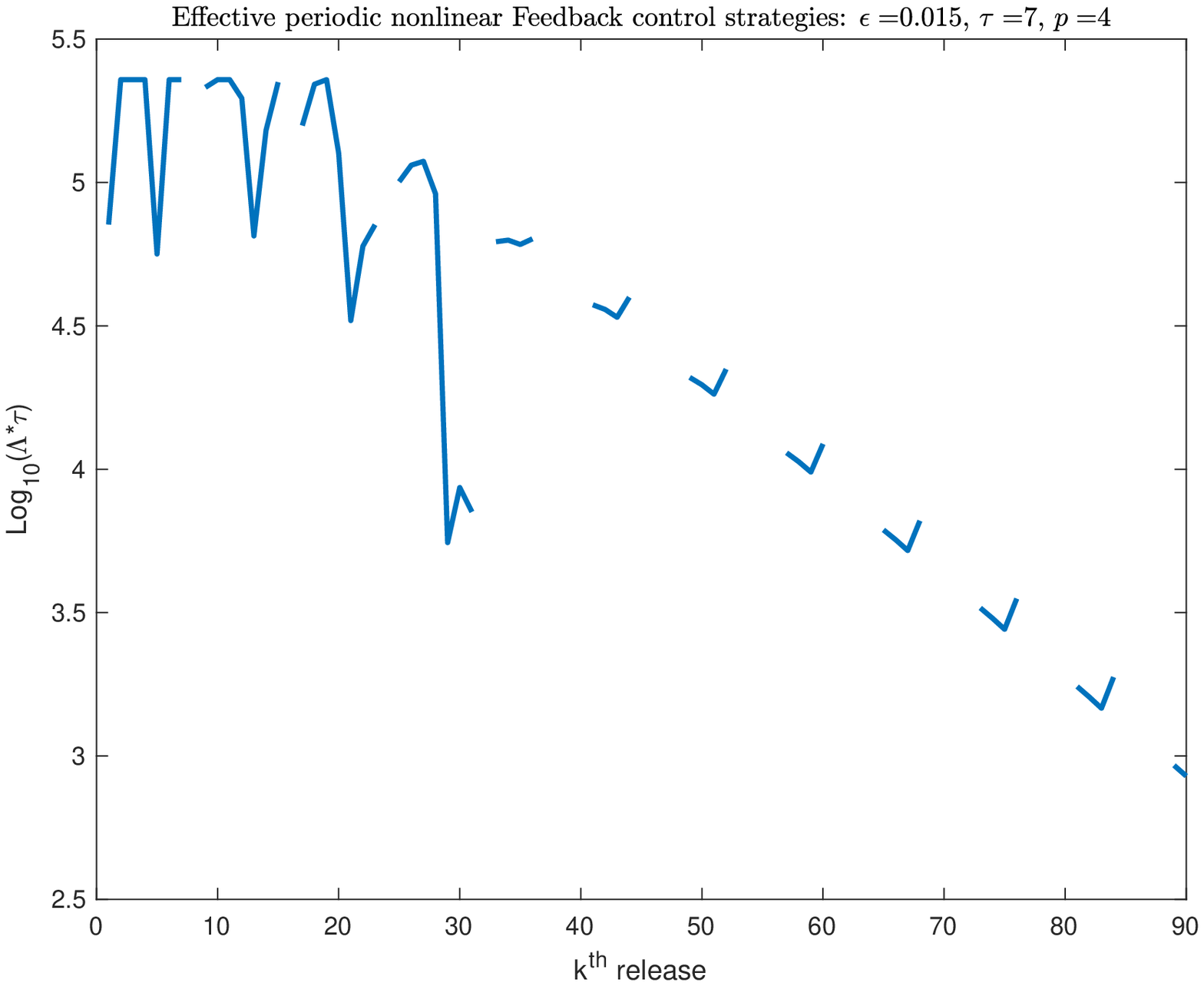} \\
 (c) & (d)
 \end{tabular}
 \caption{Residual fertility case with adulticide treatment - \textit{Aedes albopictus} - Mixed periodic impulsive  SIT control of system (\ref{eq:periodicTIS}) with $\epsilon=0.015$, $\theta=0.2 \cN_F^{-1}$, $\tau=7$ and $p=4$ - {\bf Linear} control: (a) population dynamics; (b) Field releases timing - {\bf Nonlinear} control: (c) population dynamics; (d) Field releases timing. See Tables \ref{tabPS1a} and \ref{tabPS1b}, page \pageref{tabPS1b}}
 \label{fig:1PS}
\end{figure}
For the residual fertility case, it seems that the best option is the nonlinear control with $\tau=7$, with a population estimated every $p=4$ weeks. The gain is almost $50\%$ less releases, even if we have $5$ additional releases.

As expected, the induced sterility, even low, increases not only the size of the releases but also the duration of SIT treatment, such that to stay in a realistic time experiment and release sizes, another control methods should be used, for instance one week of adulticide control before SIT starts.
\begin{table}[!th]
\centering
\begin{tabular}{|c|c|c|c|c|}   \hline
Mixed-Control & $p$ & Period (days) & Cumulative Number of  & Nb of effective Releases \\
& &  & released sterile males &  to reach  $[\bf{0},\bf{E}_1[$ \\
 \hline
Linear & $4$ & $\tau=7$  &  $1.1012722 \times 10^7$  & $49$\\
 \hline
Nonlinear  & $4$ & $\tau=7$  &  $2.596800 \times 10^6$  & $47$\\
\hline
\end{tabular}
\caption{\textit{Aedes albopictus} - Cumulative number of released sterile males and number of releases for \textbf{linear} or \textbf{nonlinear} mixed control, after one week of adulticide, , when $\theta=0.2 \cN_F^{-1}$ and $\epsilon=0.015$.}
\label{tabPS4}
\end{table}
According to Table \ref{tabPS4}, the gain is significant: around $42\%$ less insects to release than without adulticide treatment.
Thus, clearly, without adulticide or equivalent control treatment, releasing sterilized males, even with small residual fertility, is problematic and the risk of failure of the program is high. Only, a combination of controls to reduce the wild population before SIT treatment can be helpful, if the residual fertility is small enough.

\section{Conclusion}
In this work we have improved the linear feedback control developed in \cite{Bliman2019} and, in addition, we studied the possibility/risk of releasing partially sterile males, i.e. sterile males with a small, but positive, fertility rate $\epsilon$. 
A control with partial sterility is possible. However, several drawbacks occur: if the fertility is greater than $\cN_F^{-1}$, then no control is possible; if the fertility is below $\cN_F^{-1}$, then the control needs long time and (very) large releases, with the total number of released sterile males being five times more than the quantity needed in  the case of full sterility, i.e. when $\epsilon=0$ (see Table \ref{tab1b}).

Clearly, even if it is showed on a particular model, the condition $\epsilon <\cN_F^{-1}$ is always needed to guarantee that SIT works under massive releases, for almost all SIT models. However, even under that restriction, the size of the releases (or the duration of the control) can be so large that SIT alone becomes \if{practically useless}\fi unreasonable from a practical point of view. That is why a combination of control tools, including SIT, is needed \cite{Shelly2010}.

Altogether, our results highlight the importance of a very good knowledge of the pest/vector dynamics, i.e. the biological parameters and their sexual behaviors, preferably along the whole year, in order to determine the best period to start the SIT treatment. For pest/vector with a large basic offspring number, when full sterility cannot be achieved, it is clearly recommended to couple SIT with other biological control methods, like mechanical control, (pheromone, food) traps, etc.

Clearly, it seems preferable to release fully sterile males even if there is the cost in terms of fitness. Of course, this requires a sterilization protocol that insures $100\%$ sterility.

The fruit flies case, here \textit{Bactrocera dorsalis}, shows that SIT alone requires huge releases since the dynamics of the pest can be really strong. However, the model we used here does not necessarily reflects the complexity of the fruit flies dynamics, in particular their complex mating behaviors. Thus, precise models and  experiments are needed to confirm our results. However, this first insight shows that most probably a combination of control tools would be useful to better control this pest, like for instance, a combination of SIT with a Male Annihilation Technique \cite{Manoukis2019}. The results obtained for \textit{Bactrocera dorsalis} also apply to another fruit fly, \textit{Ceratitis capitata}, that may as well have a very large basic offspring number \cite{Carey1984,Pieterse2019}.

Finally, like in \cite{Dumont2012}, where an epidemiological model coupled with an SIT model was studied for the first time within the context of La R\'eunion, it would be interesting to determine whether, despite the fact that $\epsilon>\mathcal{N}_F^{-1}$, the SIT approach can be helpful to reduce the epidemiological risk, i.e. to stir $\mathcal{R}_0$ below 1 for vector-borne diseases, like chikungunya and dengue fever.

\vspace{0.5cm}
\section*{Acknowledgments}

MSA was supported by the National Council for Scientific and Technological Development (CNPq), by FAPERJ through the  {\em ``Jovem Cientista do Nosso Estado''} Program and by the  Getulio Vargas Foundation (FGV, Rio de Janeiro, Brazil) through the {\em ``Projeto de Pesquisa Aplicada''} Program.

YD acknowledges the support of the School of Applied Mathematics of FGV (FGV EMAp) that funded his visit in Rio in 2019. YD is partially supported by the ``SIT feasibility project against \textit{Aedes albopictus} in Reunion Island", TIS 2B (2020-2021), jointly funded by the French Ministry of Health and the European Regional Development Fund (ERDF). YD is (partially) supported by the DST/NRF SARChI Chair  in Mathematical Models and Methods in Biosciences and Bioengineering at the University of Pretoria (grant 82770). YD is also partially supported by the CeraTIS-Corse project, funded by the call Ecophyto 2019 (project n$^o$: 19.90.402.001), against \textit{Ceratitis capitata}. This work is done within the framework of the GEMDOTIS project (Ecophyto 2018 funding), that is ongoing in La R\'eunion. This work was also co-funded by the European Union: Agricultural Fund for Rural Development (EAFRD), by the Conseil R\'egional de La R\'eunion, the Conseil D\'epartemental de La R\'eunion, and by the Centre de Coop\'eration internationale en Recherche Agronomique pour le D\'eveloppement (CIRAD).

\end{document}